\documentclass[a4paper,twocolumn]{aa}
\usepackage{amsmath,amssymb,amsfonts}
\usepackage{widetext}
\usepackage[T1]{fontenc}
\usepackage{graphicx}
\usepackage[percent]{overpic}
\usepackage{tikz}
\usepackage{adjustbox,multirow}
\usepackage{hyperref}
\usepackage{natbib}
\bibpunct{(}{)}{;}{a}{}{,} % to follow the A&A style

\newcommand{\bb}[1]{{\boldsymbol{#1}}}
\newcommand{\hh}[1]{{\boldsymbol{\widehat #1}}}
\renewcommand{\tt}[1]{{\boldsymbol{\widetilde #1}}}
\newcommand{\tb}[1]{{\tilde{\boldsymbol #1}}}

\newcommand{\diag}{\mathrm{diag}}
\newcommand{\tr}{\mathrm{tr}}
\newcommand{\T}{\dagger}%{\intercal}

\renewcommand{\d}{\mathrm{d}}
\newcommand{\D}{\mathcal{D}}
\newcommand{\e}{\mathrm{e}}
\newcommand{\G}{\mathcal{G}}

\renewcommand{\P}{\mathcal{P}}

\newcommand{\N}{\mathbb{N}}
\newcommand{\R}{\mathbb{R}}

\begin{document}

\title{Denoising, Deconvolving, and Decomposing \\ Photon Observations}
\subtitle{Derivation of the D$^\mathsf{3}$PO Algorithm}

\titlerunning{D$^3$PO -- Denoising, Deconvolving, and Decomposing Photon Observations}

%\title{Denoising, deconvolving, and decomposing \\ photon observations}
%\subtitle{Derivation of the D$^\mathsf{3}$PO algorithm}
%
%\titlerunning{D$^3$PO -- Denoising, deconvolving, and decomposing photon observations}

\author{
    Marco~Selig\inst{\ref{inst1},\ref{inst2}}\and
    Torsten~A.~En{\ss}lin\inst{\ref{inst1},\ref{inst2}}
}

\institute{
    Max Planck Institut f\"ur Astrophysik (Karl-Schwarzschild-Stra{\ss}e~1, D-85748~Garching, Germany)\label{inst1}
    \and Ludwig-Maximilians-Universit\"at M\"unchen (Geschwister-Scholl-Platz~1, D-80539~M\"unchen, Germany) \label{inst2}
}

\date{Received 07 Nov. 2013 / Accepted DD MMM. YYYY}

\abstract{
    %% (context), aims, methods, results, (conclusions)
    The analysis of astronomical images is a non-trivial task. The D$^3$PO algorithm addresses the inference problem of denoising, deconvolving, and decomposing photon observations. Its primary goal is the simultaneous but individual reconstruction of the diffuse and point-like photon flux given a single photon count image, where the fluxes are superimposed.
    In order to discriminate between these morphologically different signal components, a probabilistic algorithm is derived in the language of information field theory based on a hierarchical Bayesian parameter model.
    The signal inference exploits prior information on the spatial correlation structure of the diffuse component and the brightness distribution of the spatially uncorrelated point-like sources.
    A maximum \emph{a~posteriori} solution and a solution minimizing the Gibbs free energy of the inference problem using variational Bayesian methods are discussed.
    Since the derivation of the solution is not dependent on the underlying position space, the implementation of the D$^3$PO algorithm uses the \textsc{NIFTy} package to ensure applicability to various spatial grids and at any resolution.
    The fidelity of the algorithm is validated by the analysis of simulated data, including a realistic high energy photon count image showing a $32 \times 32 \,\mathrm{arcmin}^2$ observation with a spatial resolution of $0.1 \,\mathrm{arcmin}$. In all tests the D$^3$PO algorithm successfully denoised, deconvolved, and decomposed the data into a diffuse and a point-like signal estimate for the respective photon flux components.
}


\maketitle

%%================================
\section{Introduction}

    An astronomical image might display multiple superimposed features, such as point sources, compact objects, diffuse emission, or background radiation. The raw photon count images delivered by high energy telescopes are far from perfect; they suffer from shot noise and distortions caused by instrumental effects. The analysis of such astronomical observations demands elaborate denoising, deconvolution, and decomposition strategies.

    The data obtained by the detection of individual photons is subject to Poissonian shot noise which is more severe for low count rates. This causes difficulty for the discrimination of faint sources against noise, and makes their detection exceptionally challenging. Furthermore, uneven or incomplete survey coverage and complex instrumental response functions leave imprints in the photon data. As a result, the data set might exhibit gaps and artificial distortions rendering the clear recognition of different features a difficult task. Point-like sources are especially afflicted by the instrument's point spread function (PSF) that smooths them out in the observed image, and therefore can cause fainter ones to vanish completely in the background noise.

    In addition to such noise and convolution effects, it is the superposition of the different objects that makes their separation ambiguous, if possible at all. In astrophysics, photon emitting objects are commonly divided into two morphological classes, diffuse sources and point sources. Diffuse sources span rather smoothly across large fractions of an image, and exhibit apparent internal correlations. Point sources, on the contrary, are local features that, if observed perfectly, would only appear in one pixel of the image. In this work, we will not distinguish between diffuse sources and background, both are diffuse contributions. Intermediate cases, which are sometimes classified as extended or compact sources, are also not considered here.

    The question arises of how to reconstruct the original source contributions, the individual signals, that caused the observed photon data. This task is an ill-posed inverse problem without a unique solution. There are a number of heuristic and probabilistic approaches that address the problem of denoising, deconvolution, and decomposition in partial or simpler settings.

    \textsc{SExtractor}~\citep{BA96} is one of the heuristic tools and the most prominent for identifying sources in astronomy. Its popularity is mostly based on its speed and easy operability. However, \textsc{SExtractor} produces a catalog of fitted sources rather than denoised and deconvolved signal estimates. \textsc{CLEAN}~\citep{H74} is commonly used in radio astronomy and attempts a deconvolution assuming there are only contributions from point sources. Therefore, diffuse emission is not optimally reconstructed in the analysis of real observations using \textsc{CLEAN}.

\onecolumn

    \begin{figure*}[!t]
        \centering
        \begin{tabular}{c}
            \begin{overpic} [scale=0.5]{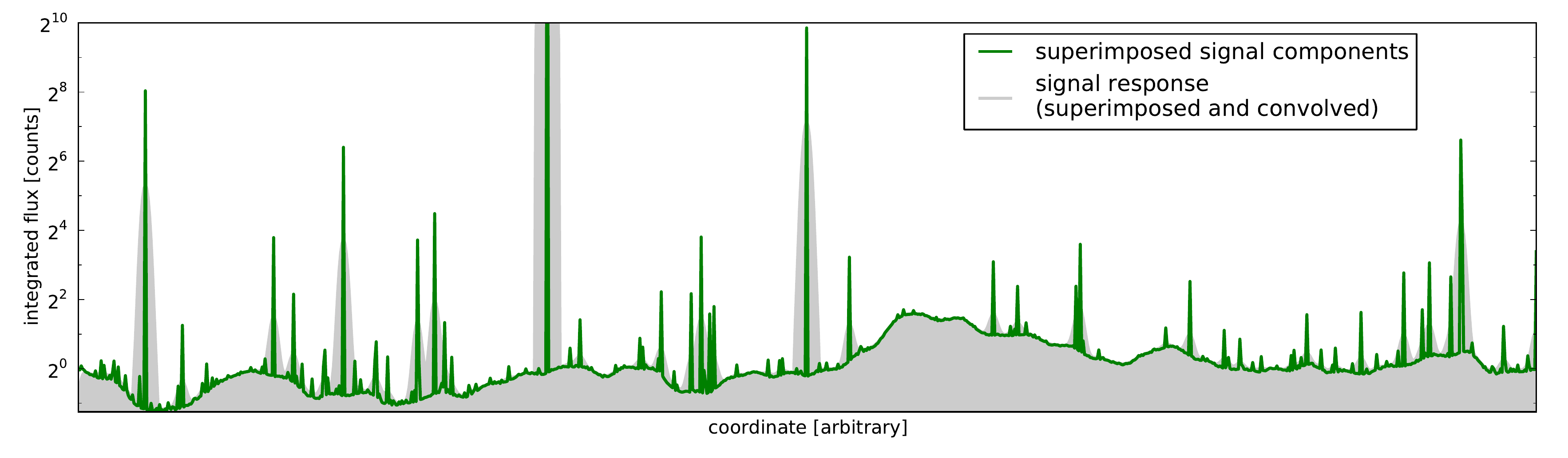} \put(-2,29){(a)} \end{overpic} \\
            \begin{overpic} [scale=0.5]{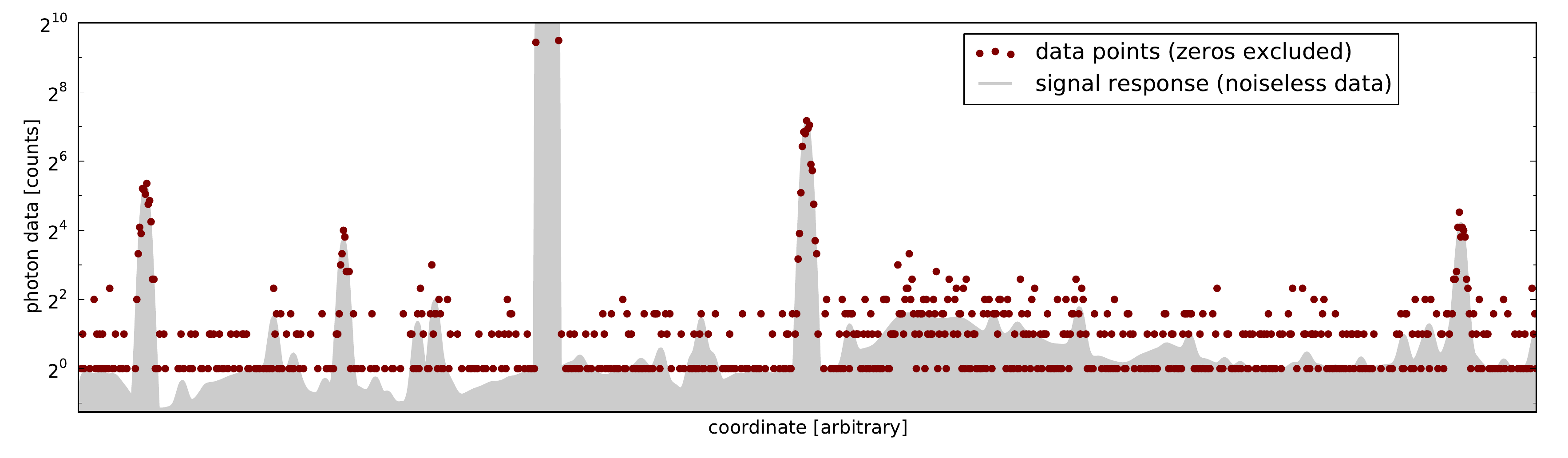} \put(-2,29){(b)} \end{overpic} \\
            \begin{overpic} [scale=0.5]{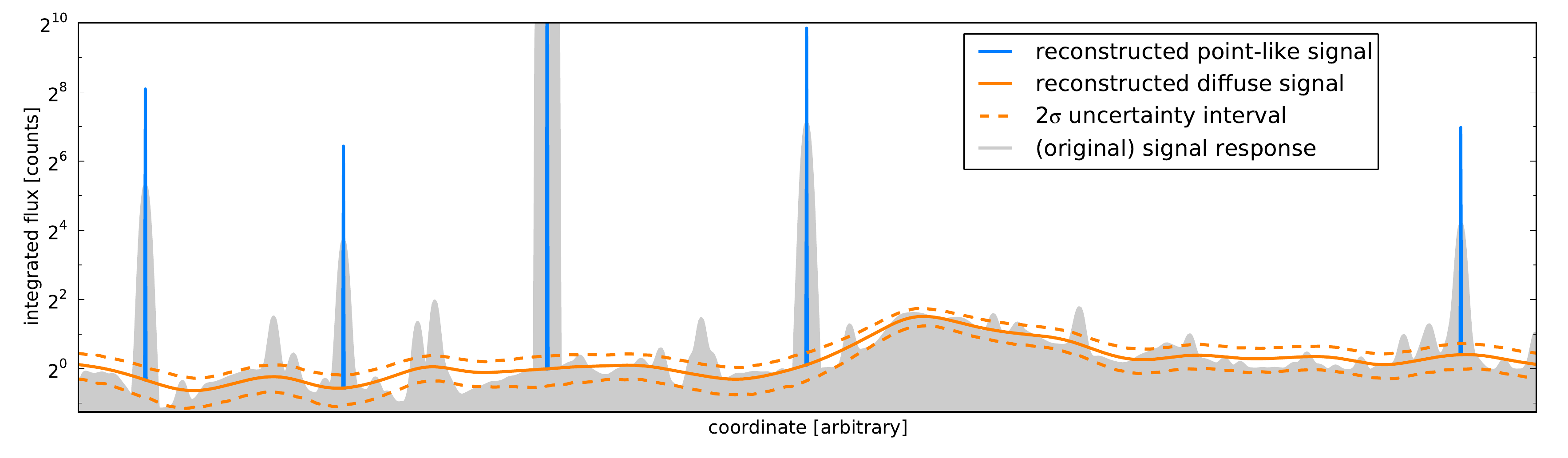} \put(-2,29){(c)} \end{overpic} \\
            \begin{overpic} [scale=0.5]{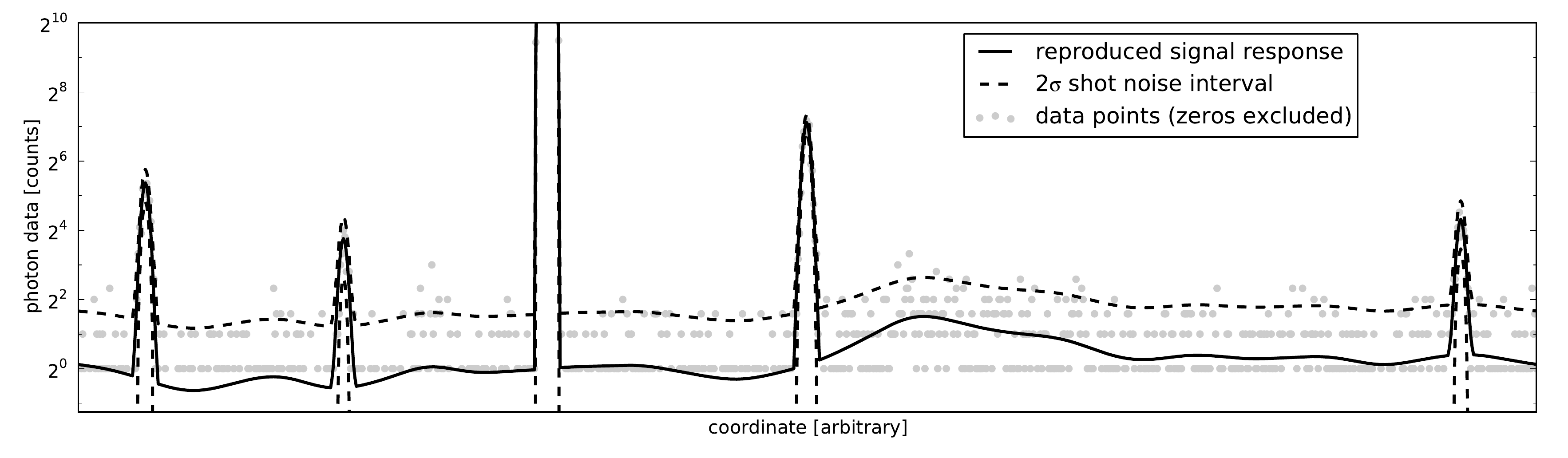} \put(-2,29){(d)} \end{overpic}
        \end{tabular}
        %\flushleft
        \caption{Illustration of a 1D reconstruction scenario with $1024$ pixels. Panel (a) shows the superimposed diffuse and point-like signal components (green solid line) and its observational response (gray contour). Panel (b) shows again the signal response representing noiseless data (gray contour) and the generated Poissonian data (red markers). Panel (c) shows the reconstruction of the point-like signal component (blue solid line), the diffuse one (orange solid line), its $2\sigma$ reconstruction uncertainty interval (orange dashed line), and again the original signal response (gray contour). The point-like signal comprises $1024$ point-sources of which only five are not too faint. Panel (d) shows the reproduced signal response representing noiseless data (black solid line), its $2\sigma$ shot noise interval (black dashed line), and again the data (gray markers).}
        \label{fig:motivation}
    \end{figure*}

\twocolumn

    \noindent
    Multiscale extensions of CLEAN \citep{C_08,RC11} improve on this, but are also not prefect \citep{JBSE13}. Decomposition techniques for diffuse backgrounds, based on the analysis of angular power spectra have recently been proposed by \citet{HPS12}.

    Inference methods, in contrast, investigate the probabilistic relation between the data and the signals. Here, the signals of interest are the different source contributions. Probabilistic approaches allow a transparent incorporation of model and \emph{a~priori} assumptions, but often result in computationally heavier algorithms.

    As an initial attempt, a maximum likelihood analysis was proposed by \citet{V82}. In later work, maximum entropy \citep{S03} and minimum $\chi^2$ methods \citep[e.g.,][]{BABR13} were applied to the INTEGRAL/SPI data reconstructing a single signal component, though.
    On the basis of sparse regularization a number of techniques exploiting waveforms \citep[based on the work by][]{H10,H11} have proven successful in performing  denoising and deconvolution tasks in different settings \citep{G+06,W07,D09,F10,D11}. For example, \citet{S10,S12} analyzed simulated (single and multi-channel) data from the Fermi $\gamma$-ray space telescope focusing on the removal of Poisson noise and deconvolution or background separation. Furthermore, a (generalized) morphological component analysis denoised, deconvolved and decomposed simulated radio data assuming Gaussian noise statistics \citep{B07,C13}.

    Still in the regime of Gaussian noise, \citet{GC08} derived a deconvolution algorithm for point and extended sources minimizing regularized least squares. They introduce an efficient convex regularization scheme at the price of \emph{a~priori} unmotivated fine tuning parameters.
    The fast algorithm PowellSnakes~I/II by \citet{CRH09,CRHL11} is capable of analyzing multi-frequency data sets and detecting point-like objects within diffuse emission regions. It relies on matched filters using PSF templates and Bayesian filters exploiting, among others, priors on source position, size, and number. PowellSnakes~II has been successfully applied to the Planck data \citep{P711}.

    The approach closest to ours is the background-source separation technique used to analyze the ROSAT data \citep{GFD09}. This Bayesian model is based on a two-component mixture model that reconstructs extended sources and (diffuse) background concurrently. The latter is, however, described by a spline model with a small number of spline sampling points.

    The strategy presented in this work aims at the simultaneous reconstruction of two signals, the diffuse and point-like photon flux. Both fluxes contribute equally to the observed photon counts, but their morphological imprints are very different. The proposed algorithm, derived in the framework of information field theory (IFT) \citep{EFK09,E13}, therefore incorporates prior assumptions in form of a hierarchical parameter model.
    The fundamentally different morphologies of diffuse and point-like contributions reflected in different prior correlations and statistics. The exploitation of these different prior models is crucial to the signal decomposition.
    In this work, we exclusively consider Poissonian noise, in particular, but not exclusively, in the low count rate regime, where the signal-to-noise ratio becomes challengingly low.
    The D$^3$PO algorithm presented here targets the simultaneous denoising, deconvolution, and decomposition of photon observations into two signals, the diffuse and point-like photon flux. This task includes the reconstruction of the harmonic power spectrum of the diffuse component from the data themselves. Moreover, the proposed algorithm provides \emph{a~posteriori} uncertainty information on both inferred signals.

    The fluxes from diffuse and point-like sources contribute equally to the observed photon counts, but their morphological imprints are very different. The proposed algorithm, derived in the framework of information field theory (IFT) \citep{EFK09,E13,E14}, therefore incorporates prior assumptions in form of a hierarchical parameter model.
    The fundamentally different morphologies of diffuse and point-like contributions reflected in different prior correlations and statistics. The exploitation of these different prior models is key to the signal decomposition.

    The diffuse and point-like signal are treated as two separate signal fields. A signal field represents an original signal appearing in nature; e.g., the physical photon flux distribution of one source component as a function of real space or sky position. In theory, a field has infinitely many degrees of freedom being defined on a continuous position space. In computational practice, however, a field needs of course to be defined on a finite grid. It is desirable that the signal field is reconstructed independently from the grid's resolution, except for potentially unresolvable features.\footnote{If the resolution of the reconstruction were increased gradually, the diffuse signal field might exhibit more and more small scale features until the information content of the given data is exhausted. From this point on, any further increase in resolution would not change the signal field reconstruction significantly. In a similar manner, the localization accuracy and number of detections of point sources might increase with the resolution until all relevant information of the data was captured. All higher resolution grids can then be regarded as acceptable representations of the continuous position space.}
    We note that the point-like signal field hosts one point source in every pixel, however, most of them might be invisibly faint. Hence, a complicated determination of the number of point sources, as many algorithms perform, is not required in our case.

    The derivation of the algorithm makes use of a wide range of Bayesian methods that are discussed below in detail with regard to their implications and applicability.
    For now, let us consider an example to emphasize the range and performance of the D$^3$PO algorithm.
    Figure~\ref{fig:motivation} illustrates a reconstruction scenario in one dimension, where the coordinate could be an angle or position (or time, or energy) in order to represent a 1D sky (or a time series, or an energy spectrum). The numerical implementation uses the \textsc{NIFTy}\footnote{\textsc{NIFTy} homepage \url{http://www.mpa-garching.mpg.de/ift/nifty/}} package \citep{S+13}. \textsc{NIFTy} permits an algorithm to be set up abstractly, independent of the finally chosen topology, dimension, or resolution of the underlying position space. In this way, a 1D prototype code can be used for development, and then just be applied in 2D, 3D, or even on the sphere.

    The remainder of this paper is structured as follows. Sec.~\ref{sec:problem} discusses the inference on photon observations; i.e., the underlying model and prior assumptions. The D$^3$PO algorithm solving this inference problem by denoising, deconvolution, and decomposition is derived in Sec.~\ref{sec:solution}. In Sec.~\ref{sec:application} the algorithm is demonstrated in a numerical application on simulated high energy photon data. We conclude in Sec.~\ref{sec:conclusion}.

%%================================
\section{Inference on photon observations}
\label{sec:problem}

%%--------------------------------
\subsection{Signal inference}
\label{sec:ift}

    Here, a signal is defined as an unknown quantity of interest that we want to learn about. The most important information source on a signal is the data obtained in an observation to measure the signal.
    Inferring a signal from an observational data set poses a fundamental problem because of the presence of noise in the data and the ambiguity that several possible signals could have produced the same data, even in the case of negligible noise.

    For example, given some image data like photon counts, we want to infer the underlying photon flux distribution. This physical flux is a continuous scalar field that varies with respect to time, energy, and observational position. The measured photon count data, however, is restricted by its spatial and energy binning, as well as its limitations in energy range and observation time. Basically, all data sets are finite for practical reasons, and therefore cannot capture all of the infinitely many degrees of freedom of the underlying continuous signal field.

    There is no exact solution to such signal inference problems, since there might be (infinitely) many signal field configurations that could lead to the same data. This is why a probabilistic data analysis, which does not pretend to calculate the correct field configuration but provides expectation values and uncertainties of the signal field, is appropriate for signal inference.

    Given a data set $\bb{d}$, the \emph{a~posteriori} probability distribution $P(\bb{s}|\bb{d})$ judges how likely a potential signal $\bb{s}$ is. This posterior is given by Bayes' theorem,
    \begin{align}
        P(\bb{s}|\bb{d}) &= \frac{P(\bb{d}|\bb{s}) \, P(\bb{s})}{P(\bb{d})}
        , \label{eq:bayes}
    \end{align}
    as a combination of the likelihood $P(\bb{d}|\bb{s})$, the signal prior $P(\bb{s})$, and the evidence $P(\bb{d})$, which serves as a normalization. The likelihood characterizes how likely it is to measure data set $\bb{d}$ from a given signal field $\bb{s}$. It covers all processes that are relevant for the measurement of $\bb{d}$. The prior describes the knowledge about $\bb{s}$ without considering the data, and should, in general, be less restrictive than the likelihood.

    IFT is a Bayesian framework for the inference of signal fields exploiting mathematical methods for theoretical physics. A signal field, $\bb{s} = s(x)$, is a function of a continuous position $x$ in some position space $\Omega$. In order to avoid a dependence of the reconstruction on the partition of $\Omega$, the according calculus regarding fields is geared to preserve the continuum limit \citep{E13,E14,S+13}.
    In general, we are interested in the \emph{a~posteriori} mean estimate $\bb{m}$ of the signal field given the data, and its (uncertainty) covariance $\bb{D}$, defined as
    \begin{align}
        \bb{m} &= \left< \bb{s} \right>_{(\bb{s}|\bb{d})} = \int\D\bb{s} \; \bb{s} \; P(\bb{s}|\bb{d})
        , \label{eq:map} \\
        \bb{D} &= \left< (\bb{m}-\bb{s})(\bb{m}-\bb{s})^\T \right>_{(\bb{s}|\bb{d})}
        , \label{eq:error}
    \end{align}
    where $\T$ denotes adjunction and $\left< \,\cdot\, \right>_{(\bb{s}|\bb{d})}$ the expectation value with respect to the posterior probability distribution $P(\bb{s}|\bb{d})$.\footnote{This expectation value is computed by a path integral, $\int \D\bb{s}$, over the complete phase space of the signal field $\bb{s}$; i.e., all possible field configurations.}

    In the following, the posterior of the physical photon flux distribution of two morphologically different source components given a data set of photon counts is build up piece by piece according to Eq.~\eqref{eq:bayes}.

%%--------------------------------
\subsection{Poissonian likelihood}
\label{sec:likelihood}

    The images provided by astronomical high energy telescopes typically consist of integer photon counts that are binned spatially into pixels. Let $d_i$ be the number of detected photons, also called events, in pixel $i$, where $i \in \{1,\dots,N_\mathrm{pix}\} \subset \N$.

    The kind of signal field we would like to infer from such data is the causative photon flux distribution. The photon flux, $\bb{\rho} = \rho(x)$, is defined for each position $x$ on the observational space $\Omega$. In astrophysics, this space $\Omega$ is typically the $\mathcal{S}^2$ sphere representing an all-sky view, or a region within $\R^2$ representing an approximately plane patch of the sky. %The abstract postion space could also include dimensions for energy and time, however, in this work merely spatial dimensions are considered.
    The flux $\bb{\rho}$ might express different morphological features, which can be classified into a diffuse and point-like component. The exact definitions of the diffuse and point-like flux should be specified \emph{a~priori}, without knowledge of the data, and are addressed in Sec.~\ref{sec:s_prior} and \ref{sec:u_prior}, respectively. At this point it shall suffices to say that the diffuse flux varies smoothly on large spatial scales, while the flux originating from point sources is fairly local.
    These two flux components are superimposed,
    \begin{align}
        \bb{\rho} &= \bb{\rho}_\mathrm{diffuse} + \bb{\rho}_\mathrm{point-like} = \rho_0 \left( \e^\bb{s} + \e^\bb{u} \right)
        , \label{eq:superposition}
    \end{align}
    where we introduced the dimensionless diffuse and point-like signal fields, $\bb{s}$ and $\bb{u}$, and the constant $\rho_0$ which absorbs the physical dimensions of the photon flux; i.e., events per area per energy and time interval. The exponential function in Eq.~\eqref{eq:superposition} is applied componentwise. In this way, we naturally account for the strict positivity of the photon flux at the price of a non-linear change of variables, from the flux to its natural logarithm.

    A measurement apparatus observing the photon flux $\bb{\rho}$ is expected to detect a certain number of photons $\bb{\lambda}$. This process can be modeled by a linear response operator $\bb{R}_0$ as follows,
    \begin{align}
        \bb{\lambda} &= \bb{R}_0 \bb{\rho} = \bb{R} \left( \e^\bb{s} + \e^\bb{u} \right)
        , \label{eq:l}
    \end{align}
    where $\bb{R} = \bb{R}_0 \rho_0$. This reads for pixel $i$,
    \begin{align}
        \lambda_i &= \int_\Omega \d x \; R_i(x) \left( \e^{s(x)} + \e^{u(x)} \right)
        . \label{eq:l_i}
    \end{align}
    The response operator $\bb{R}_0$ comprises all aspects of the measurement process; i.e., all instrument response functions. This includes the survey coverage, which describes the instrument's overall exposure to the observational area, and the instrument's PSF, which describes how a point source is imaged by the instrument.

    The superposition of different components and the transition from continuous coordinates to some discrete pixelization, cf. Eq.~\eqref{eq:l_i}, cause a severe loss of information about the original signal fields. In addition to that, measurement noise distorts the signal's imprint in the data. The individual photon counts per pixel can be assumed to follow a Poisson distribution $\P$ each. Therefore, the likelihood of the data $\bb{d}$ given an expected number of events $\bb{\lambda}$ is modeled as a product of statistically independent Poisson processes,
    \begin{align}
        P(\bb{d}|\bb{\lambda}) &= \prod_i \P(d_i,\lambda_i) = \prod_i \frac{1}{d_i!} \; \lambda_i^{d_i} \e^{-\lambda_i}
        . \label{eq:likelihood}
    \end{align}
    The Poisson distribution has a signal-to-noise ratio of $\sqrt{\bb{\lambda}}$ which scales with the expected number of photon counts. Therefore, Poissonian shot noise is most severe in regions with low photon fluxes. This makes the detection of faint sources in high energy astronomy a particularly challenging task, as  X- and $\gamma$-ray photons are sparse.

    The likelihood of photon count data given a two component photon flux is hence described by the Eqs.~\eqref{eq:l} and \eqref{eq:likelihood}. Rewriting this likelihood $P(\bb{d}|\bb{s},\bb{u})$ in form of its negative logarithm yields the information Hamiltonian $H(\bb{d}|\bb{s},\bb{u})$,\footnote{Throughout this work we define $H(\,\cdot\,) = - \log P(\,\cdot\,)$, and absorb constant terms into a normalization constant $H_0$ in favor of clarity.}
    \begin{align}
        H(\bb{d}|\bb{s},\bb{u}) &= -\log P(\bb{d}|\bb{s},\bb{u})
        \\
        &= H_0 + \bb{1}^\T \bb{\lambda} - \bb{d}^\T \log(\bb{\lambda})
        \\
        &= H_0 + \bb{1}^\T \bb{R} \left( \e^\bb{s} + \e^\bb{u} \right) - \bb{d}^\T \log \left( \bb{R} \left( \e^\bb{s} + \e^\bb{u} \right) \right)
        , \label{eq:H_likelihood}
    \end{align}
    where the ground state energy $H_0$ comprises all terms constant in $\bb{s}$ and $\bb{u}$, and $\bb{1}$ is a constant data vector being $1$ everywhere.

%%--------------------------------
\subsection{Prior assumptions}
\label{sec:prior}

    The diffuse and point-like signal fields, $\bb{s}$ and $\bb{u}$, contribute equally to the likelihood defined by Eq.~\eqref{eq:H_likelihood}, and thus leaving it completely degenerate. On the mere basis of the likelihood, the full data set could be explained by the diffuse signal alone, or only by point-sources, or any other conceivable combination. In order to downweight intuitively implausible solutions, we introduce priors.
    The priors discussed in the following address the morphology of the different photon flux contributions, and define diffuse and point-like in the first place. These priors aid the reconstruction by providing some remedy for the degeneracy of the likelihood.
    The likelihood describes noise and convolution properties, and the prior describe the individual morphological properties. Therefore, the denoising and deconvolution of the data towards the total photon flux $\bb{\rho}$ is primarily likelihood driven, but for a decomposition of the total photon flux into $\bb{\rho}^{(s)}$ and $\bb{\rho}^{(u)}$, the signal priors are imperative.

%% - - - - - - - - - - - - - - - -
\subsubsection{Diffuse component}
\label{sec:s_prior}

    The diffuse photon flux, $\bb{\rho}^{(s)} = \rho_0 \e^\bb{s}$, is strictly positive and might vary in intensity over several orders of magnitude. Its morphology shows cloudy patches with smooth fluctuations across spatial scales; i.e., one expects similar values of the diffuse flux in neighboring locations. In other words, the diffuse component exhibits spatial correlations. A log-normal model for $\bb{\rho}^{(s)}$ satisfies those requirements according to the maximum entropy principle \citep{OSBE12,K13}.
    If the diffuse photon flux follows a multivariate log-normal distribution, the diffuse signal field $\bb{s}$ obeys a multivariate Gaussian distribution $\G$,
    \begin{align}
        P(\bb{s}|\bb{S}) &= \G(\bb{s},\bb{S}) = \frac{1}{\sqrt{\det[2\pi\bb{S}]}} \; \exp\left( - \frac{1}{2} \bb{s}^\T \bb{S}^{-1} \bb{s} \right)
        , \label{eq:s_prior}
    \end{align}
    with a given covariance $\bb{S} = \left< \bb{s} \bb{s}^\T \right>_{(\bb{s}|\bb{S})}$. This covariance describes the strength of the spatial correlations, and thus the smoothness of the fluctuations.

    A convenient parametrization of the covariance $\bb{S}$ can be found, if the signal field $\bb{s}$ is \emph{a~priori} not known to distinguish any position or orientation axis; i.e., its correlations only depend on relative distances. This is equivalent to assume $\bb{s}$ to be statistically homogeneous and isotropic. Under this assumption, $\bb{S}$ is diagonal in the harmonic basis\footnote{The basis in which the Laplace operator is diagonal is denoted harmonic basis. If $\Omega$ is a $n$-dimensional Euclidean space $\R^n$ or Torus $\mathcal{T}^n$, the harmonic basis is the Fourier basis; if $\Omega$ is the $\mathcal{S}^2$ sphere, the harmonic basis is the spherical harmonics basis.} of the position space $\Omega$ such that
    \begin{align}
        \bb{S} &= \sum_k \e^{\tau_k} \bb{S}_k
        ,
    \end{align}
    where $\tau_k$ are spectral parameters and $\bb{S}_k$ are projections onto a set of disjoint harmonic subspaces of $\Omega$. These subspaces are commonly denoted as spectral bands or harmonic modes. The set of spectral parameters, $\bb{\tau} = \{\tau_k\}_k$, is then the logarithmic power spectrum of the diffuse signal field $\bb{s}$ with respect to the chosen harmonic basis denoted by $k$.

    However, the diffuse signal covariance is in general unknown \emph{a~priori}. This requires the introduction of another prior for the covariance, or for the set of parameters $\bb{\tau}$ describing it adequately. This approach of hyperpriors on prior parameters creates a hierarchical parameter model.

%% - - - - - - - - - - - - - - - -
\subsubsection{Unknown power spectrum}
\label{sec:p_prior}

    The lack of knowledge of the power spectrum, requires its reconstruction from the same data the signal is inferred from \citep{WLL04,JKWE10,EF11,J13}. Therefore, two \emph{a~priori} constraints for the spectral parameters $\bb{\tau}$, which describe the logarithmic power spectrum, are incorporated in the model.

    The power spectrum is unknown and might span over several orders of magnitude. This implies a logarithmically uniform prior for each element of the power spectrum, and a uniform prior for each spectral parameter $\tau_k$, respectively. We initially assume independent inverse-Gamma distributions $\mathcal{I}$ for the individual elements,
    \begin{align}
        P(\e^{\bb{\tau}}|\bb{\alpha},\bb{q}) &= \prod_k \mathcal{I}(\e^{\tau_k},\alpha_k,q_k)
        \\
        &= \prod_k \frac{q_k^{\alpha_k-1}}{\Gamma(\alpha_k-1)} \; \e^{-\left(\alpha_k \tau_k + q_k \e^{-\tau_k} \right)}
        ,
    \end{align}
    and hence
    \begin{align}
        P_\mathrm{un}(\bb{\tau}|\bb{\alpha},\bb{q}) &= \prod_k \mathcal{I}(\e^{\tau_k},\alpha_k,q_k) \left| \frac{\d\e^{\tau_k}}{\d\tau_k} \right|
        \\
        &\propto \exp\Big( - (\bb{\alpha} - \bb{1})^\T \bb{\tau} - \bb{q}^\T \e^\bb{-\tau} \Big)
        ,
    \end{align}
    where $\bb{\alpha} = \{\alpha_k\}_k$ and $\bb{q} = \{q_k\}_k$ are the shape and scale parameters, and $\Gamma$ denotes the Gamma function. In the limit of $\alpha_k \rightarrow 1$ and $q_k \rightarrow 0 \; \forall k$, the inverse-Gamma distributions become asymptotically flat on a logarithmic scale, and thus $P_\mathrm{un}$ constant.\footnote{If $P(\tau_k = \log z) = \mathrm{const.}$, then a substitution yields $P(z) = P(\log z) \; |\d(\log z)/\d z| \propto z^{-1} \sim \mathcal{I}(z,\alpha \rightarrow 1, q \rightarrow 0)$.} Small non-zero scale parameters, $0 < q_k$, provide lower limits for the power spectrum that, in practice, lead to more stable inference algorithms.

    So far, the variability of the individual elements of the power spectrum is accounted for, but the question about their correlations has not been addressed. Empirically, power spectra of a diffuse signal field do not exhibit wild fluctuation or change drastically over neighboring modes. They rather show some sort of spectral smoothness. Moreover, for diffuse signal fields that were shaped by local and causal processes, we might expect a finite correlation support in position space. This translates into a smooth power spectrum.
    In order to incorporate spectral smoothness, we employ a prior introduced by \citet{EF11,OSBE12}. This prior is based on the second logarithmic derivative of the spectral parameters $\bb{\tau}$, and favors power spectra that obey a power law. It reads
    \begin{align}
        P_\mathrm{sm}(\bb{\tau}|\bb{\sigma}) &\propto \exp\left( - \frac{1}{2} \bb{\tau}^\T \bb{T} \bb{\tau} \right)
        ,
    \end{align}
    with
    \begin{align}
        \bb{\tau}^\T \bb{T} \bb{\tau} &= \int\d(\log k) \frac{1}{\sigma_k^2} \left( \frac{\partial^2 \tau_k}{\partial (\log k)^2} \right)^2
        ,
    \end{align}
    where $\bb{\sigma} = \{\sigma_k\}_k$ are Gaussian standard deviations specifying the tolerance against deviation from a power-law behavior of the power spectrum.
    A choice of $\sigma_k = 1 \; \forall k$ would typically allow for a change in the power law's slope of $1$ per e-fold in $k$.
    In the limit of $\sigma_k \rightarrow \infty \; \forall k$, no smoothness is enforced upon the power spectrum.

    The resulting prior for the spectral parameters is given by the product of the priors discussed above,
    \begin{align}
        P(\bb{\tau}|\bb{\alpha},\bb{q},\bb{\sigma}) &= P_\mathrm{un}(\bb{\tau}|\bb{\alpha},\bb{q}) \; P_\mathrm{sm}(\bb{\tau}|\bb{\sigma})
        . \label{eq:t_prior}
    \end{align}
    The parameters $\bb{\alpha},\bb{q}$ and $\bb{\sigma}$ are considered to be given as part of the hierarchical Bayesian model, and provide a flexible handle to model our knowledge on the scaling and smoothness of the power spectrum.

%% - - - - - - - - - - - - - - - -
\subsubsection{Point-like component}
\label{sec:u_prior}

    The point-like photon flux, $\bb{\rho}^{(u)} = \rho_0 \e^\bb{u}$, is supposed to originate from very distant astrophysical sources. These sources appear morphologically point-like to an observer because their actual extent is negligible given the extreme distances. This renders point sources to be spatially local phenomena. The photon flux contributions of neighboring point sources can (to zeroth order approximation) be assumed to be statistically independent of each other. Even if the two sources are very close on the observational plane, their physical distance might be huge.
    Even in practice, the spatial cross-correlation of point sources is negligible. Therefore, statistically independent priors for the photon flux contribution of each point-source are assumed in the following.

    Because of the spatial locality of a point source, the corresponding photon flux signal is supposed to be confined to a single spot, too. If the point-like signal field, defined over a continuous position space $\Omega$, is discretized properly\footnote{The numerical discretization of information fields is described in great detail in \citet{S+13}.}, this spot is sufficiently identified by an image pixel in the reconstruction. A discretization, $\rho(x \in \Omega) \rightarrow (\rho_x)_x$, is an inevitable step since the algorithm is to be implemented in a computer environment anyway. Nevertheless, we have to ensure that the \emph{a~priori} assumptions do not depend on the chosen discretization but satisfy the continuous limit.

    Therefore, the prior for the point-like signal component factorizes spatially,
    \begin{align}
        P(\bb{\rho}^{(u)}) &= \prod_x P(\rho_x^{(u)})
        , \label{eq:prod}
    \end{align}
    but the functional form of the priors are yet to be determined. This model allows the point-like signal field to host one point source in every pixel. Most of these point sources are expected to be invisibly faint contributing negligibly to the total photon flux. However, the point sources which are just identifiable from the data are pinpointed in the reconstruction. In this approach, there is no necessity for a complicated determination of the number and position of sources.

    For the construction of a prior, that the photon flux is a strictly positive quantity also needs to be considered. Thus, a simple exponential prior,
    \begin{align}
        P(\rho_x^{(u)}) &\propto \exp\left( - \rho_x^{(u)}/\rho_0 \right)
        ,
    \end{align}
    has been suggested \citep[e.g.,][]{GFD09}. It has the advantage of being (easily) analytically treatable, but its physical implications are questionable. This distribution strongly suppresses high photon fluxes in favor of lower ones. The maximum entropy prior, which is also often applied, is even worse because it corresponds to a brightness distribution,\footnote{The so-called maximum entropy regularization $\sum_x (\rho_x^{(u)}/\rho_0) \log(\rho_x^{(u)}/\rho_0)$ of the log-likelihood can be regarded as log-prior, cf. Eqs.~\eqref{eq:prod} and \eqref{eq:ME}.}
    \begin{align}
        P(\rho_x^{(u)}) &\propto \left( \rho_x^{(u)}/\rho_0 \right)^{\left( - \rho_x^{(u)}/\rho_0 \right)}
        . \label{eq:ME}
    \end{align}
    The following (rather crude) consideration might motivate a more astrophysical prior. Say the universe hosts a homogeneous distribution of point sources. The number of point sources would therefore scale with the observable volume; i.e., with distance cubed. Their apparent brightness, which is reduced because of the spreading of the light rays; i.e., a proportionality to the distance squared. Consequently, a power-law behavior between the number of point sources and their brightness with a slope $\beta = \tfrac{3}{2}$ is to be expected \citep{F68,MH11}. However, such a plain power law diverges at $0$, and is not necessarily normalizable. Furthermore, Galactic and extragalactic sources cannot be found in arbitrary distances owing to the finite size of the Galaxy and the cosmic (past) light cone. Imposing an exponential cut-off above $0$ onto the power law yields an inverse-Gamma distribution, which has been shown to be an appropriate prior for point-like photon fluxes \citep{GFD09,CRH09,CRHL11}.

    The prior for the point-like signal field is therefore derived from a product of independent inverse-Gamma distributions,\footnote{A possible extension of this prior model that includes spatial correlations would be an inverse-Wishart distribution for $\diag[ \bb{\rho}^{(u)} ]$.}
    \begin{align}
        P(\bb{\rho}^{(u)}|\bb{\beta},\bb{\eta}) &= \prod_x \mathcal{I}(\rho_x^{(u)},\beta_x,\rho_0\eta_x)
        \\
        &= \prod_x \frac{(\rho_0\eta_x)^{\beta_x-1}}{\Gamma(\beta_x-1)} \left( \rho_x^{(u)} \right)^{-\beta_x} \exp\left( - \frac{\rho_0\eta_x}{\rho_x^{(u)}} \right)
        ,
    \end{align}
    yielding
    \begin{align}
        P(\bb{u}|\bb{\beta},\bb{\eta}) &= \prod_x \mathcal{I}(\rho_0\e^{u_x},\beta_k,\rho_0\eta_k) \left| \frac{\d\rho_0\e^{u_x}}{\d u_x} \right|
        \\
        &\propto \exp\Big( - (\bb{\beta} - \bb{1})^\T \bb{u} - \bb{\eta}^\T \e^\bb{-\bb{u}} \Big)
        , \label{eq:u_prior}
    \end{align}
    where $\bb{\beta} = \{\beta_x\}_x$ and $\bb{\eta} = \{\eta_x\}_x$ are the shape and scale parameters. The latter is responsible for the cut-off of vanishing fluxes, and should be chosen adequately small in analogy to the spectral scale parameters $\bb{q}$. The determination of the shape parameters is more difficile. The geometrical argument above suggests a universal shape parameter, $\beta_x = \tfrac{3}{2} \; \forall x$. A second argument for this value results from demanding \emph{a~priori} independence of the discretization. If we choose a coarser resolution that would add up the flux from two point sources at merged pixels, then our prior should still be applicable. The universal value of $\tfrac{3}{2}$ indeed fulfills this requirement as shown in Appendix~\ref{app:stacking}. There it is also shown that $\eta$ has to be chosen resolution dependent, though.

%%--------------------------------
\subsection{Parameter model}
\label{sec:parameter_model}

\begin{figure}[t!]
        \centering
        \begin{tikzpicture}
            [c/.style={circle,minimum size=2em,text centered,thin},
             r/.style={rectangle,minimum size=2em,text centered,thin},
             v/.style={->,shorten >=1pt,>=stealth,thick}]
            \node(a)at(-2,5)[r,text width=3em,draw]{$\alpha,\;q$};
            \node(z)at(0,5)[r,draw]{$\sigma$};
            \node(b)at(2,5)[r,text width=3em,draw]{$\beta,\;\eta$};
            \node(t)at(-1,4)[c,draw]{$\bb{\tau}$};
            \node(s)at(-1,3)[c,draw]{$\bb{s}$};
            \node(u)at(2,3)[c,draw]{$\bb{u}$};
            \node(g)at(0.5,2)[c,draw,dashed]{$\bb{\rho}$};
            \node(l)at(0.5,1)[c,draw,dashed]{$\bb{\lambda}$};
            \node(d)at(0.5,0)[r,draw]{$\bb{d}$};
            \draw[v](a.south)--(t);
            \draw[v](z.south)--(t);
            \draw[v](t)--(s);
            \draw[v](b)--(u);
            \draw[v](s)--(g);
            \draw[v](u)--(g);
            \draw[v](g)--(l);
            \draw[v](l)--(d);
        \end{tikzpicture}
        \flushleft
        \caption{Graphical model of the model parameters $\alpha$, $q$, $\sigma$, $\beta$, and $\eta$, the logarithmic spectral parameters $\bb{\tau}$, the diffuse signal field $\bb{s}$, the point-like signal field $\bb{u}$, the total photon flux $\bb{\rho}$, the expected number of photons $\bb{\lambda}$, and the observed photon count data $\bb{d}$.}
        \label{fig:hierarchy}
    \end{figure}
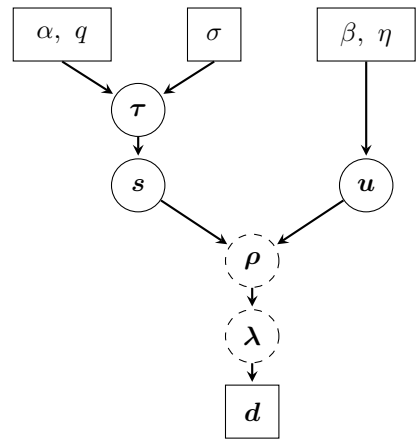

    Figure~\ref{fig:hierarchy} gives an overview of the parameter hierarchy of the suggested Bayesian model. The data $\bb{d}$ is given, and the diffuse signal field $\bb{s}$ and the point-like signal field $\bb{u}$ shall be reconstructed from that data. The logarithmic power spectrum $\bb{\tau}$ is a set of nuisance parameters that also need to be reconstructed from the data in order to accurately model the diffuse flux contributions. The model parameters form the top layer of this hierarchy and are given to the reconstruction algorithm. This set of model parameters can be boiled down to five scalars, namely $\alpha$, $q$, $\sigma$, $\beta$, and $\eta$, if one defines $\bb{\alpha} = \alpha \bb{1}$, etc. The incorporation of the scalars in the inference is possible in theory, but this would increase the computational complexity dramatically.

    We discussed reasonable values for these scalars to be chosen \emph{a~priori}. If additional information sources, such as theoretical power spectra or object catalogs, are available the model parameters can be adjusted accordingly. In Sec.~\ref{sec:application}, different parameter choices for the analysis of simulated data are investigated.

%%================================
\section{Denoising, deconvolution, and decomposition}
\label{sec:solution}

    The likelihood model, describing the measurement process, and the prior assumptions for the signal fields and the power spectrum of the diffuse component yield a well-defined inference problem. The corresponding posterior is given by
    \begin{align}
        P(\bb{s},\bb{\tau},\bb{u}|\bb{d}) &= \frac{P(\bb{d}|\bb{s},\bb{u})\; P(\bb{s}|\bb{\tau}) \; P(\bb{\tau}|\alpha,q,\sigma) \; P(\bb{u}|\beta,\eta)}{P(\bb{d})}
        , \label{eq:posterior}
    \end{align}
    which is a complex form of Bayes' theorem~\eqref{eq:bayes}.

    Ideally, we would now calculate the \emph{a~posteriori} expectation values and uncertainties according to Eqs.~\eqref{eq:map} and \eqref{eq:error} for the diffuse and point-like signal fields, $\bb{s}$ and $\bb{u}$, as well as for the logarithmic spectral parameters $\bb{\tau}$. However, an analytical evaluation of these expectation values is not possible because of the complexity of the posterior.

    The posterior is non-linear in the signal fields and, except for artificially constructed data, non-convex. It, however, is more flexible and therefore allows for a more comprehensive description of the parameters to be inferred \citep{KGV83,GG84}.

    Numerical approaches involving Markov chain Monte Carlo methods \citep{MU49,M+53} are possible, but hardly feasible because of the huge parameter phase space. Nevertheless, similar problems have been addressed by elaborate sampling techniques \citep{WLL04,JKWE10,JK10,J13}.

    Here, two approximative algorithms with lower computational costs are derived. The first one uses the maximum \emph{a~posteriori} (MAP) approximation, the second one minimizes the Gibbs free energy of an approximate posterior ansatz in the spirit of variational Bayesian methods. The fidelity and accuracy of these two algorithms are compared in a numerical application in Sec.~\ref{sec:application}.

%%--------------------------------
\subsection{Posterior maximum}
\label{sec:MAP}

    The posterior maximum and mean coincide, if the posterior distribution is symmetric and single peaked. In practice, this often holds -- at least in good approximation --, so that the maximum \emph{a~posteriori} approach can provide suitable estimators.
    This can either be achieved using a $\delta$-distribution at the posterior's mode,
    \begin{align}
        \left< \bb{s} \right>_{(\bb{s}|\bb{d})} &\overset{\mathrm{MAP}\text{-}\delta}{\approx} \int\D\bb{s} \; \bb{s} \; \delta(\bb{s} - \bb{s}_\mathrm{mode})
        , \label{eq:MAPmap}
    \end{align}
    or using a Gaussian approximation around this point,
    \begin{align}
        \left< \bb{s} \right>_{(\bb{s}|\bb{d})} &\overset{\mathrm{MAP}\text{-}\G}{\approx} \int\D\bb{s} \; \bb{s} \; \G(\bb{s} - \bb{s}_\mathrm{mode},\bb{D}_\mathrm{mode})
        , \label{eq:MAPmap2}
    \end{align}
    Both approximations require us to find the mode, which is done by extremizing the posterior.

    Instead of the complex posterior distribution, it is convenient to consider the information Hamiltonian, defined by its negative logarithm,
    \begin{align}
        H(\bb{s},\bb{\tau},\bb{u}|\bb{d}) &= - \log P(\bb{s},\bb{\tau},\bb{u}|\bb{d})
        \\
        &= H_0 + \bb{1}^\T \bb{R} \left( \e^\bb{s} + \e^\bb{u} \right) - \bb{d}^\T \log \left( \bb{R} \left( \e^\bb{s} + \e^\bb{u} \right) \right)
        \notag \\
        &\quad + \frac{1}{2} \log\left( \det\left[ \bb{S} \right] \right) + \frac{1}{2} \bb{s}^\T \bb{S}^{-1} \bb{s}
        \label{eq:H} \\
        &\quad + (\bb{\alpha} - \bb{1})^\T \bb{\tau} + \bb{q}^\T \e^\bb{-\tau} + \frac{1}{2} \bb{\tau}^\T \bb{T} \bb{\tau}
        \notag \\
        &\quad + (\bb{\beta} - \bb{1})^\T \bb{u} + \bb{\eta}^\T \e^{-\bb{u}}
        \notag
        ,
    \end{align}
    where all terms constant in $\bb{s}$, $\bb{\tau}$, and $\bb{u}$ have been absorbed into a ground state energy $H_0$, cf. Eqs.~\eqref{eq:likelihood}, \eqref{eq:s_prior}, \eqref{eq:t_prior}, and \eqref{eq:u_prior}, respectively.

    The MAP solution, which maximizes the posterior, minimizes the Hamiltonian.
    This minimum can thus be found by taking the first (functional) derivatives of the Hamiltonian with respect to $\bb{s}$, $\bb{\tau}$, and $\bb{u}$ and equating them with zero. Unfortunately, this yields a set of implicit, self-consistent equations rather than an explicit solution. However, these equations can be solved by an iterative minimization of the Hamiltonian using a steepest descent method for example, see Sec.~\ref{sec:algorithm} for details.

    In order to better understand the structure of the MAP solution, we consider the minimum $(\bb{s},\bb{\tau},\bb{u}) = (\bb{m}^{(s)},\bb{\tau}^\star,\bb{m}^{(u)})$. The resulting filter formulas for the diffuse and point-like signal field read
    \begin{align}
        \frac{\partial H}{\partial\bb{s}} \bigg|_\mathrm{min} &= \bb{0} = \left( \bb{1} - \bb{d}/\bb{l} \right)^\T \bb{R} \ast \e^{\bb{m}^{(s)}} + {\bb{S}^\star}^{-1} \bb{m}^{(s)}
        , \label{eq:s_MAP} \\
        \frac{\partial H}{\partial\bb{u}} \bigg|_\mathrm{min} &= \bb{0} = \left( \bb{1} - \bb{d}/\bb{l} \right)^\T \bb{R} \ast \e^{\bb{m}^{(u)}} + \bb{\beta} - \bb{1} - \bb{\eta} \ast \e^{-\bb{m}^{(u)}}
        , \label{eq:u_MAP}
    \end{align}
    with
    \begin{align}
        \bb{l} &= \bb{R} \left( \e^{\bb{m}^{(s)}} + \e^{\bb{m}^{(u)}} \right)
        , \label{eq:l_MAP} \\
        \bb{S}^\star &= \sum_k \e^{\tau_k^\star} \bb{S}_k
        .
    \end{align}
    Here, $\ast$ and $/$ denote componentwise multiplication and division, respectively. The first term in Eq.~\eqref{eq:s_MAP} and \eqref{eq:u_MAP}, which comes from the likelihood, vanishes in case $\bb{l} = \bb{d}$.
    We note that $\bb{l} = \bb{\lambda}|_\mathrm{min}$ describes the most likely number of photon counts, not the expected number of photon counts $\bb{\lambda} = \left< \bb{d} \right>_{(\bb{d}|\bb{s},\bb{u})}$, cf. Eqs.~\eqref{eq:l} and \eqref{eq:likelihood}.
    Disregarding the regularization by the priors, the solution would overfit; i.e., noise features are partly assigned to the signal fields in order to achieve an unnecessarily close agreement with the data. However, the \emph{a~priori} regularization suppresses this tendency to some extend.

    The second derivative of the Hamiltonian describes the curvature around the minimum, and therefore approximates the (inverse) uncertainty covariance,
    \begin{align}
        \frac{\partial^2 H}{\partial\bb{s}\partial\bb{s}^\T} \bigg|_\mathrm{min} &\approx {\bb{D}^{(s)}}^{-1}
        , &
        \frac{\partial^2 H}{\partial\bb{u}\partial\bb{u}^\T} \bigg|_\mathrm{min} &\approx {\bb{D}^{(u)}}^{-1}
        . \label{eq:D_MAP}
    \end{align}
    The closed form of $\bb{D}^{(s)}$ and $\bb{D}^{(u)}$ is given explicitly in Appendix~\ref{app:covariance}.

    The filter formula for the power spectrum, which is derived from a first derivative of the Hamiltonian with respect to $\bb{\tau}$, yields
    \begin{align}
        \e^{\bb{\tau}^\star} &= \frac{\bb{q} + \frac{1}{2} \left( \tr\left[ \bb{m}^{(s)}{\bb{m}^{(s)}}^\T \bb{S}_k^{-1} \right] \right)_k}{\bb{\gamma} + \bb{T} \bb{\tau}^\star}
        , \label{eq:t_MAP}
    \end{align}
    where $\bb{\gamma} = (\bb{\alpha} - \bb{1}) + \frac{1}{2} \left( \tr\left[ \bb{S}_k {\bb{S}_k}^{-1} \right] \right)_k$. This formula is in accordance with the results by \citet{EF11,OSBE12}. It has been shown by the former authors that such a filter exhibits a perception threshold; i.e., on scales where the signal-response-to-noise ratio drops below a certain bound the reconstructed signal power becomes vanishingly low. This threshold can be cured by a better capture of the \emph{a~posteriori} uncertainty structure.

%%--------------------------------
\subsection{Posterior approximation}
\label{sec:Gibbs}

    In order to overcome the analytical infeasibility as well as the perception threshold, we seek an approximation to the true posterior.
    Instead of approximating the expectation values of the posterior, approximate posteriors are investigated in this section. In case the approximation is good, the expectation values of the approximate posterior should then be close to the real ones.

    The posterior given by Eq.~\eqref{eq:posterior} is inaccessible because of the entanglement of the diffuse signal field $\bb{s}$, its logarithmic power spectrum $\bb{\tau}$, and the point-like signal field $\bb{u}$. The involvement of $\bb{\tau}$ can been simplified by a mean field approximation,
    \begin{align}
        P(\bb{s},\bb{\tau},\bb{u}|\bb{d}) &\approx Q = Q_s(\bb{s},\bb{u}|\bb{\mu},\bb{d}) \;  Q_\tau(\bb{\tau}|\bb{\mu},\bb{d})
        , \label{eq:PQ}
    \end{align}
    where $\bb{\mu}$ denotes an abstract mean field mediating some information between the signal field tuple $(\bb{s},\bb{u})$ and $\bb{\tau}$ that are separated by the product ansatz in Eq.~\eqref{eq:PQ}.
    This mean field is fully determined by the problem, as it represents effective (rather than additional) degrees of freedom. It is only needed implicitly for the derivation, an explicit formula can be found in Appendix~\ref{app:about_t}, though.

    Since the \emph{a~posteriori} mean estimates for the signal fields and their uncertainty covariances are of primary interest, a Gaussian approximation for $Q_s$ that accounts for correlation between $\bb{s}$ and $\bb{u}$ would be sufficient. Hence, our previous approximation is extended by setting
    \begin{align}
        Q_s(\bb{s},\bb{u}|\bb{\mu},\bb{d}) = \G(\bb{\varphi},\bb{D})
        ,
    \end{align}
    with
    \begin{align}
        \bb{\varphi} &= \begin{pmatrix} \bb{s} - \bb{m}^{(s)} \\ \bb{u} - \bb{m}^{(u)} \end{pmatrix}
        , &
        \bb{D} &= \left( \begin{array}{ll} \bb{D}^{(s)} & \bb{D}^{(su)} \\ {\bb{D}^{(su)}}^\T & \bb{D}^{(u)} \end{array} \right)
        . \label{eq:gau}
    \end{align}
    This Gaussian approximation is also a convenient choice in terms of computational complexity because of its simple analytic structure.

    The goodness of the approximation $P \approx Q$ can be quantified by an information theoretical measure, see Appendix~\ref{app:KLG}. The Gibbs free energy of the inference problem,
    \begin{align}
        G &= \big< H(\bb{s},\bb{\tau},\bb{u}|\bb{d}) \big>_Q - \big< - \log Q(\bb{s},\bb{\tau},\bb{u}|\bb{d}) \big>_Q
        ,
    \end{align}
    which is equivalent to the Kullback-Leibler divergence $D_\mathrm{KL}(Q,P)$, is chosen as such a measure \citep{EW10}.

    In favor of comprehensibility, we suppose the solution for the logarithmic power spectrum $\bb{\tau}^\star$ is known for the moment. The Gibbs free energy is then calculated by plugging in the Hamiltonian, and evaluating the expectation values\footnote{The second likelihood term in Eq.~\eqref{eq:G}, $\bb{d}^\T \log(\bb{\lambda})$, is thereby expanded according to
    \begin{align}
        \log(x) &= \log\left<x\right> - \sum_{\nu=2}^\infty \frac{(-1)^\nu}{\nu} \left< \left( \frac{x}{\left<x\right>} - 1 \right)^\nu \right>
        \notag \\
        &\approx \log\left<x\right> + \mathcal{O}\left( \left< x^2 \right> \right)
        , \notag
    \end{align}
    under the assumption $x \approx \left<x\right>$.},
    \begin{align}
        G &= G_0 + \big< H(\bb{s},\bb{u}|\bb{d}) \big>_{Q_s} - \frac{1}{2} \log\left( \det\left[ \bb{D} \right] \right)
        \\
        &= G_1 + \bb{1}^\T \bb{l} - \bb{d}^\T \left\{ \log(\bb{l}) - \sum_{\nu=2}^\infty \frac{(-1)^\nu}{\nu} \big< \left( \bb{\lambda} / \bb{l} - 1 \right)^\nu \big>_{Q_s} \right\}
        \notag \\
        &\quad + \frac{1}{2} {\bb{m}^{(s)}}^\T {\bb{S}^\star}^{-1} \bb{m}^{(s)} + \frac{1}{2} \tr\left[ \bb{D}^{(s)} {\bb{S}^\star}^{-1} \right]
        \label{eq:G} \\
        &\quad + (\bb{\beta} - \bb{1})^\T \bb{m}^{(u)} + \bb{\eta}^\T \e^{-\bb{m}^{(u)} + \tfrac{1}{2} \hh{D}^{(u)}}
        \notag \\
        &\quad - \frac{1}{2} \log\left( \det\left[ \bb{D} \right] \right)
        , \notag
    \end{align}
    with
    \begin{align}
        \bb{\lambda} &= \bb{R} \left( \e^\bb{s} + \e^\bb{u} \right)
        , \\
        \bb{l} &= \left< \bb{\lambda}\right>_{Q_s} = \bb{R} \left( \e^{\bb{m}^{(s)} + \tfrac{1}{2} \hh{D}^{(s)}} + \e^{\bb{m}^{(u)} + \tfrac{1}{2} \hh{D}^{(u)}} \right)
        , \label{eq:l_Gibbs} \\
        \bb{S}^\star &= \sum_k \e^{\tau_k^\star} \bb{S}_k
        , \mathrm{\ and} \\
        \hh{D} &= \diag\left[ \bb{D} \right]
        .
    \end{align}
    Here, $G_0$ and $G_1$ carry all terms independent of $\bb{s}$ and $\bb{u}$. In comparison to the Hamiltonian given in Eq.~\eqref{eq:H}, there are a number of correction terms that now also consider the uncertainty covariances of the signal estimates properly. For example, the expectation values of the photon fluxes differ comparing $\bb{l}$ in Eq.~\eqref{eq:l_MAP} and \eqref{eq:l_Gibbs} where it now describes the expectation value of $\bb{\lambda}$ over the approximate posterior. In case $\bb{l} = \bb{\lambda}$ the explicit sum in Eq.~\eqref{eq:G} vanishes. Since this sum includes powers of $\left< \bb{\lambda}^{\nu > 2} \right>_{Q_s}$ its evaluation would require all entries of $\bb{D}$ to be known explicitly. In order to keep the algorithm computationally feasible, this sum shall hereafter be neglected. This is equivalent to truncating the corresponding expansion at second order; i.e., $\nu = 2$. It can be shown that, in consequence of this approximation, the cross-correlation $\bb{D}^{(su)}$ equals zero, and $\bb{D}$ becomes block diagonal.

    Without these second order terms, the Gibbs free energy reads
    \begin{align}
        G &= G_1 + \bb{1}^\T \bb{l} - \bb{d}^\T \log(\bb{l})
        \notag \\
        &\quad + \frac{1}{2} {\bb{m}^{(s)}}^\T {\bb{S}^\star}^{-1} \bb{m}^{(s)} + \frac{1}{2} \tr\left[ \bb{D}^{(s)} {\bb{S}^\star}^{-1} \right]
        \\
        &\quad + (\bb{\beta} - \bb{1})^\T \bb{m}^{(u)} + \bb{\eta}^\T \e^{-\bb{m}^{(u)} + \tfrac{1}{2} \hh{D}^{(u)}}
        \notag \\
        &\quad - \frac{1}{2} \log\left( \det\left[ \bb{D}^{(s)} \right] \right) - \frac{1}{2} \log\left( \det\left[ \bb{D}^{(u)} \right] \right)
        . \notag
    \end{align}
    Minimizing the Gibbs free energy with respect to $\bb{m}^{(s)}$, $\bb{m}^{(u)}$, $\bb{D}^{(s)}$, and $\bb{D}^{(u)}$ would optimize the fitness of the posterior approximation $P \approx Q$.
    Filter formulas for the Gibbs solution can be derived by taking the derivative of $G$ with respect to the approximate mean estimates,
    \begin{align}
        \frac{\partial G}{\partial\bb{m}^{(s)}} = \bb{0} &= \left( \bb{1} - \bb{d}/\bb{l} \right)^\T \bb{R} \ast \e^{\bb{m}^{(s)} + \tfrac{1}{2} \hh{D}^{(s)}} + {\bb{S}^\star}^{-1} \bb{m}^{(s)}
        , \label{eq:s_Gibbs} \\
        \frac{\partial G}{\partial\bb{m}^{(u)}} = \bb{0} &= \left( \bb{1} - \bb{d}/\bb{l} \right)^\T \bb{R} \ast \e^{\bb{m}^{(u)} + \tfrac{1}{2} \hh{D}^{(u)}}
        \label{eq:u_Gibbs}\\
        &\quad + \bb{\beta} - \bb{1} - \bb{\eta} \ast \e^{-\bb{m}^{(u)} + \tfrac{1}{2} \hh{D}^{(u)}}
        , \notag
    \end{align}
    This filter formulas again account for the uncertainty of the mean estimates in comparison to the MAP filter formulas in Eq.~\eqref{eq:s_MAP} and \eqref{eq:u_MAP}. The uncertainty covariances can be constructed by either taking the second derivatives,
    \begin{align}
        \frac{\partial^2 G}{\partial\bb{m}^{(s)}\partial{\bb{m}^{(s)}}^\T} &\approx {\bb{D}^{(s)}}^{-1}
        , &
        \frac{\partial^2 G}{\partial\bb{m}^{(u)}\partial{\bb{m}^{(u)}}^\T} &\approx {\bb{D}^{(u)}}^{-1}
        , \label{eq:D_Gibbs}
    \end{align}
    or setting the first derivatives of $G$ with respect to the uncertainty covariances equal to zero matrices,
    \begin{align}
        \frac{\partial G}{\partial D_{xy}^{(s)}} &= 0
        , &
        \frac{\partial G}{\partial D_{xy}^{(u)}} &= 0
        . \label{eq:D_Gibbs_2}
    \end{align}
    The closed form of $\bb{D}^{(s)}$ and $\bb{D}^{(u)}$ is given explicitly in Appendix~\ref{app:covariance}.

    So far, the logarithmic power spectrum $\bb{\tau}^\star$, and with it $\bb{S}^\star$, have been supposed to be known. The mean field approximation in Eq.~\eqref{eq:PQ} does not specify the approximate posterior $Q_\tau(\bb{\tau}|\bb{\mu},\bb{d})$, but it can be retrieved by variational Bayesian methods \citep{JGJS99,WW13}, according to the procedure detailed in Appendix~\ref{app:variation}. The subsequent Appendix~\ref{app:about_t} discusses the derivation of an solution for $\bb{\tau}$ by extremizing $Q_\tau$. This result, which was also derived in \citet{OSBE12}, applies to the inference problem discussed here, yielding
    \begin{align}
        \e^{\bb{\tau}^\star} &= \frac{\bb{q} + \frac{1}{2} \left( \tr\left[ \left( \bb{m}^{(s)}{\bb{m}^{(s)}}^\T + \bb{D}^{(s)} \right) \bb{S}_k^{-1} \right] \right)_k}{\bb{\gamma} + \bb{T} \bb{\tau}^\star}
        . \label{eq:t_Gibbs}
    \end{align}
    Again, this solution includes a correction term in comparison to the MAP solution in Eq.~\eqref{eq:t_MAP}. Since $\bb{D}^{(s)}$ is positive definite, it contributes positive to the (logarithmic) power spectrum, and therefore reduces the possible perception threshold further.

    We note that this is a minimal Gibbs free energy solution that maximizes $Q_\tau$. A proper calculation of $\left< \bb{\tau} \right>_{Q_\tau}$ might include further correction terms, but their derivation is not possible in closed form. Moreover, the above used diffuse signal covariance ${\bb{S}^\star}^{-1}$ should be replaced by $\left< \bb{S}^{-1} \right>_{Q_\tau}$ adding further correction terms to the filter formulas.

    In order to keep the computational complexity on a feasible level, all these higher order corrections are not considered here. The detailed characterization of their implications and implementation difficulties is left for future investigation.

%%--------------------------------
\subsection{Physical flux solution}

    To perform calculations on the logarithmic fluxes is convenient for numerical reasons, but it is the physical fluxes that are actually of interest to us.
    Given the chosen approximation, we can compute the posterior expectation values of the diffuse and point-like photon flux, $\bb{\rho}^{(s)}$ and $\bb{\rho}^{(u)}$, straight forwardly,
    \begin{align}
        \left< \bb{\rho}^{(\,\cdot\,)} \right>_P
        \overset{\mathrm{MAP}\text{-}\delta}{\approx} \left< \bb{\rho}^{(\,\cdot\,)} \right>_\delta &= \rho_0 \e^{\bb{m}_\mathrm{mode}^{(\,\cdot\,)}}
        , \label{eq:r_MAP} \\
        \overset{\mathrm{MAP}\text{-}\G}{\approx} \left< \bb{\rho}^{(\,\cdot\,)} \right>_\G &= \rho_0 \e^{\bb{m}_\mathrm{mode}^{(\,\cdot\,)} + \tfrac{1}{2} \hh{D}_\mathrm{mode}^{(\,\cdot\,)}}
        , \label{eq:r_MAP2} \\
        \overset{\mathrm{Gibbs}}{\approx} \left< \bb{\rho}^{(\,\cdot\,)} \right>_Q &= \rho_0 \e^{\bb{m}_\mathrm{mean}^{(\,\cdot\,)} + \tfrac{1}{2} \hh{D}_\mathrm{mean}^{(\,\cdot\,)}}
        , \label{eq:r_Gibbs}
    \end{align}
    in accordance with Eq.~\eqref{eq:MAPmap}, \eqref{eq:MAPmap2}, or \eqref{eq:PQ}, respectively. Those solutions differ from each other in terms of the involvement of the posterior's mode or mean, and in terms of the inclusion of the uncertainty information, see subscripts.

    In general, the mode approximation holds for symmetric, single peaked distributions, but can perform poorly in other cases \citep[e.g.,][]{EF11}. The exact form of the posterior considered here is highly complex because of the many degrees of freedom. In a dimensionally reduced frame, however, the posterior appears single peaked and exhibits a negative skewness.\footnote{For example, the posterior $P(s|d)$ for a one-dimensional diffuse signal is proportional to $\exp(-\tfrac{1}{2} s^2 + d s - \exp(s))$, whereby all other parameters are fixed to unity. Analogously, $P(u|d) \propto \exp(d u - 2\,\mathrm{cosh}(u))$.}
    Although this is not necessarily generalizable, it suggest a superiority of the posterior mean compared to the MAP because of the asymmetry of the distribution.
    Nevertheless, the MAP approach is computationally cheaper compared to the Gibbs approach that requires permanent knowledge of the uncertainty covariance.

    The uncertainty of the reconstructed photon flux can be approximated as for an ordinary log-normal distribution,
    \begin{align}
        \left< {\bb{\rho}^{(\,\cdot\,)}}^2 \right>_P - \left< \bb{\rho}^{(\,\cdot\,)} \right>_P^2
        \overset{\mathrm{MAP}}{\approx} \left< \bb{\rho}^{(\,\cdot\,)} \right>_\G^2 &\left( \e^{\hh{D}_\mathrm{mode}^{(\,\cdot\,)}} - 1 \right)
        , \label{eq:var_MAP} \\
        \overset{\mathrm{Gibbs}}{\approx} \left< \bb{\rho}^{(\,\cdot\,)} \right>_Q^2 &\left( \e^{\hh{D}_\mathrm{mean}^{(\,\cdot\,)}} - 1 \right)
        , \label{eq:var_Gibbs}
    \end{align}
    where the square root of the latter term would describe the relative uncertainty.

%%--------------------------------
\subsection{Imaging algorithm}
\label{sec:algorithm}

    \begin{figure*}[t]
        \centering
        \begin{tabular}{ccc}
            \begin{overpic} [scale=0.35]{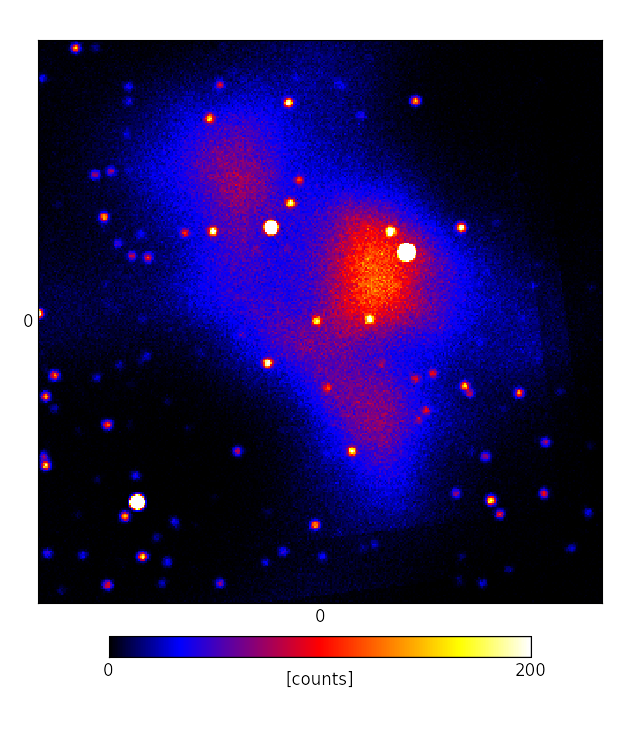}       \put(-3,97){(a)} \end{overpic} &
            \begin{overpic} [scale=0.35]{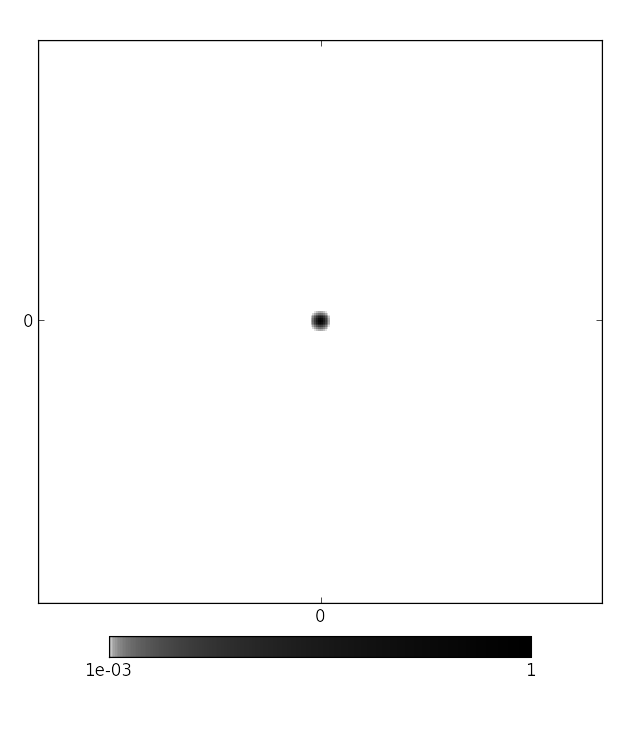}      \put(-3,97){(b)} \end{overpic} &
            \begin{overpic} [scale=0.35]{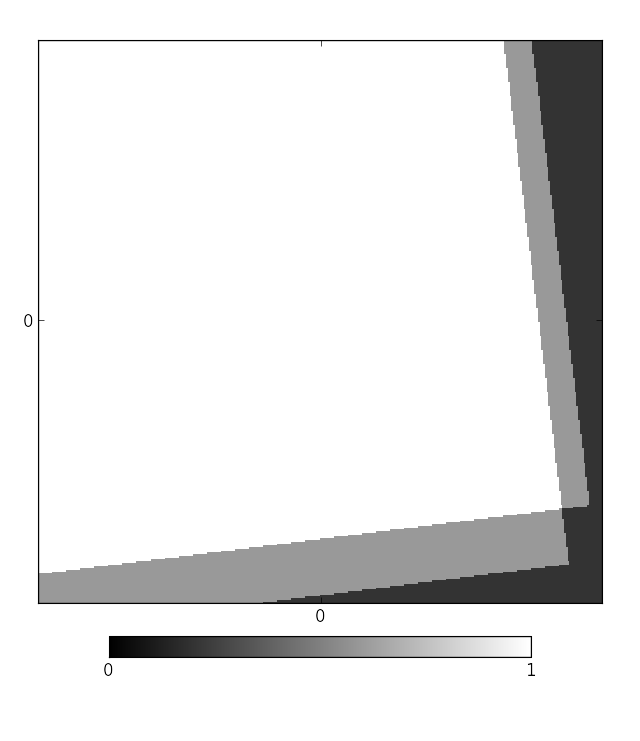}      \put(-3,97){(c)} \end{overpic} \\
            \begin{overpic} [scale=0.35]{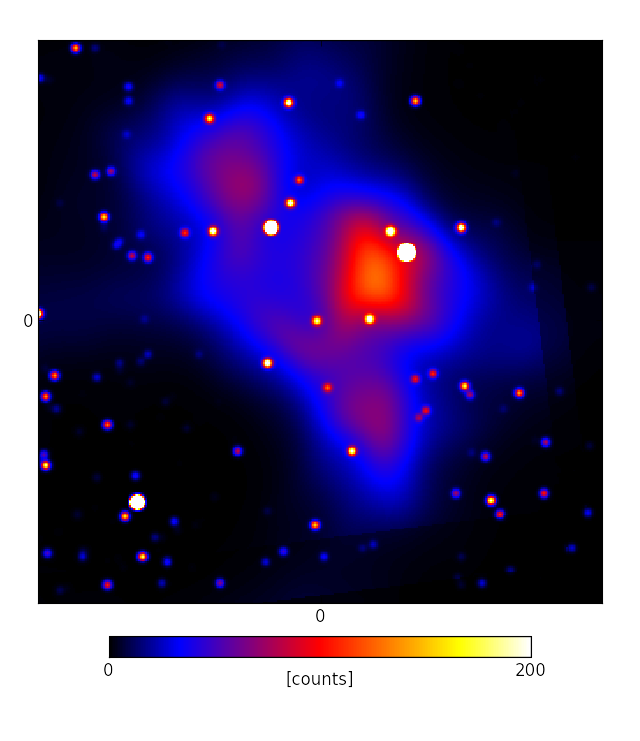}   \put(-3,97){(d)} \end{overpic} &
            \begin{overpic} [scale=0.35]{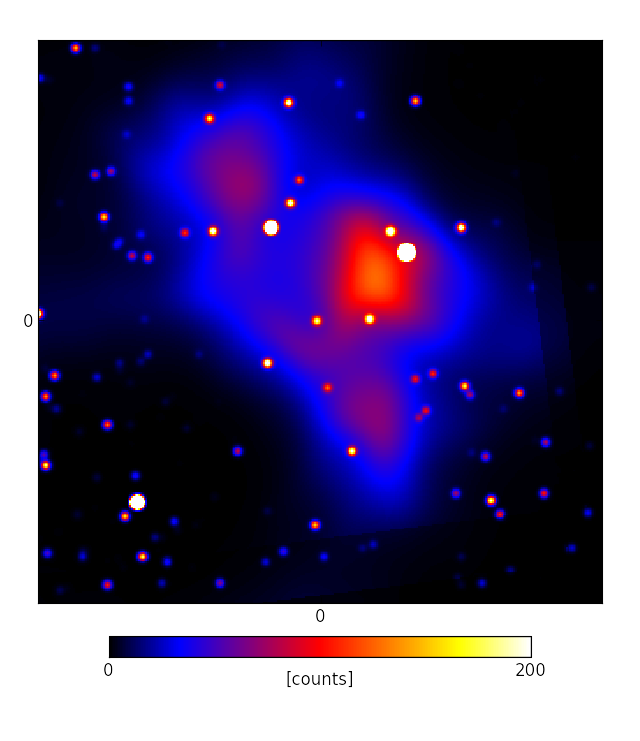}  \put(-3,97){(e)} \end{overpic} &
            \begin{overpic} [scale=0.35]{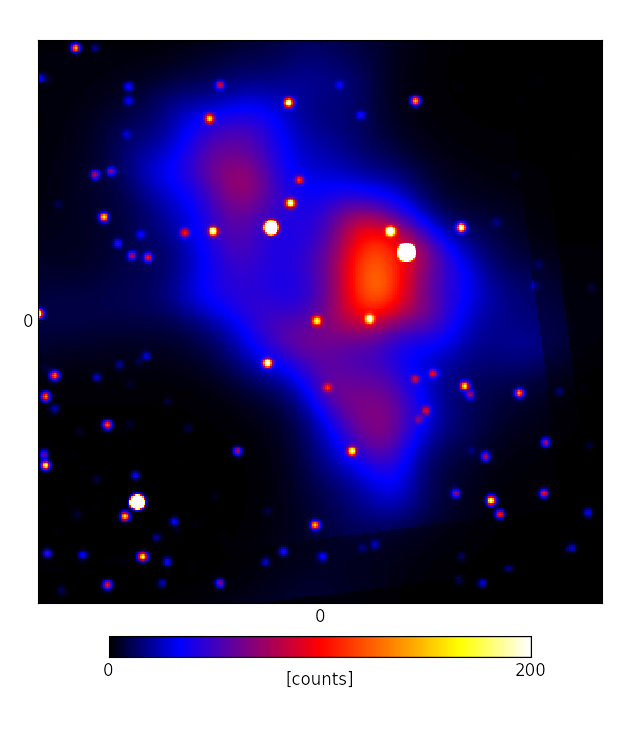} \put(-3,97){(f)} \end{overpic} \\
        \end{tabular}
        \flushleft
        \caption{Illustration of the data and noiseless, but reconvolved, signal responses of the reconstructions. Panel (a) shows the data from a mock observation of a $32 \times 32 \,\mathrm{arcmin}^2$ patch of the sky with a resolution of $0.1 \,\mathrm{arcmin}$ corresponding to a total of $102\,400$ pixels. The data had been convolved with a Gaussian-like PSF (FWHM $\approx 0.2 \,\mathrm{arcmin}$ $= 2$ pixels, finite support of $1.1 \,\mathrm{arcmin}$ $= 11$ pixels) and masked because of an uneven exposure. Panel (b) shows the centered convolution kernel. Panel (c) shows the exposure mask.
        The bottom panels show the reconvolved signal response $\bb{R}\left<\bb{\rho}\right>$ of a reconstruction using a different approach each, namely (d) MAP-$\delta$, (e) MAP-$\G$, and (f) Gibbs. All reconstructions shown here and in the following figures used the same model parameters: $\alpha=1$, $q=10^{-12}$, $\sigma=10$, $\beta = \tfrac{3}{2}$, and $\eta = 10^{-4}$.}
        \label{fig:d}
    \end{figure*}

    \begin{figure*}[t]
        \centering
        \begin{tabular}{ccc}
            \begin{overpic} [scale=0.35]{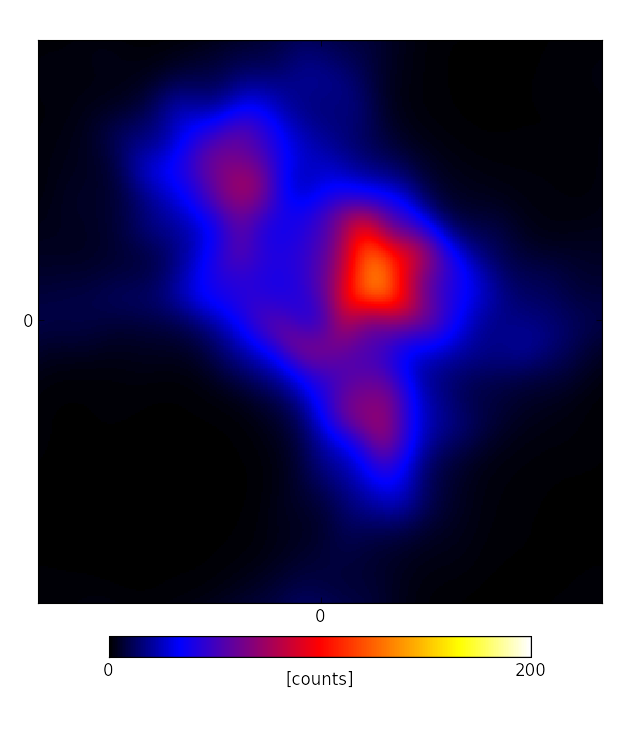}    \put(-3,97){(a)} \end{overpic} &
            \begin{overpic} [scale=0.35]{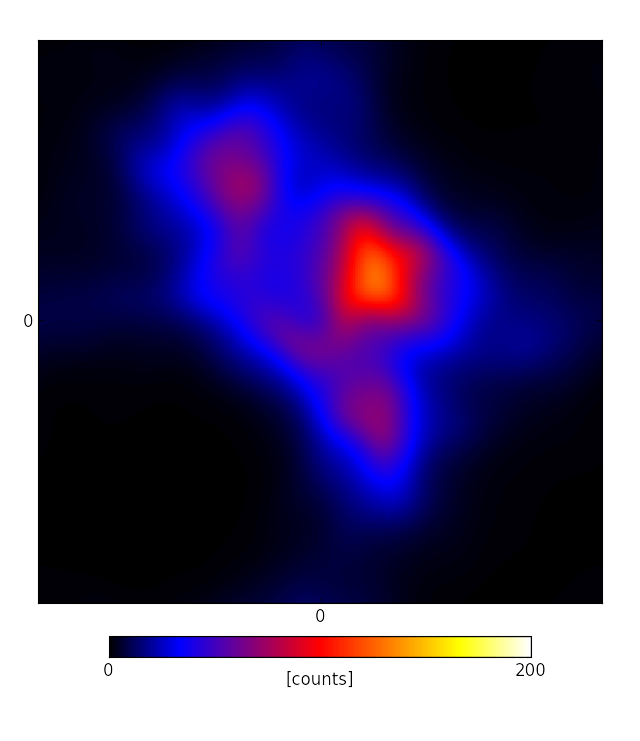}   \put(-3,97){(b)} \end{overpic} &
            \begin{overpic} [scale=0.35]{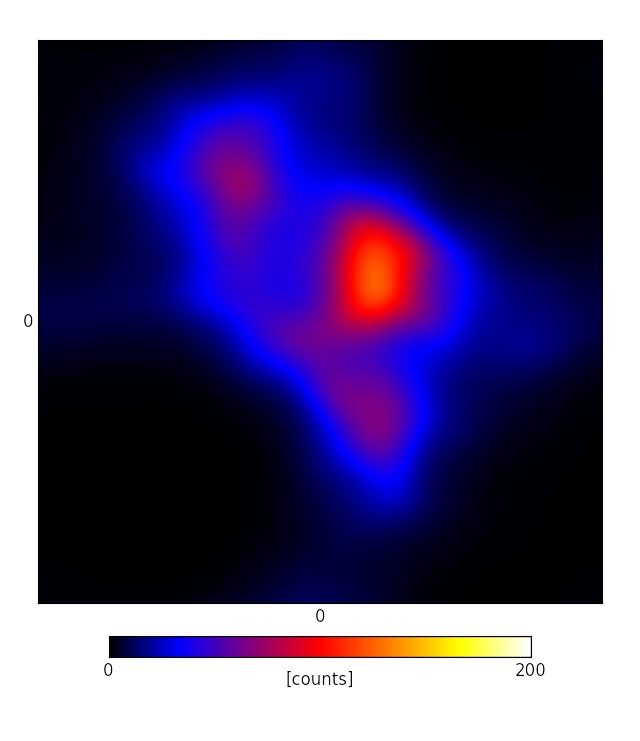}  \put(-3,97){(c)} \end{overpic} \\
            \begin{overpic} [scale=0.35]{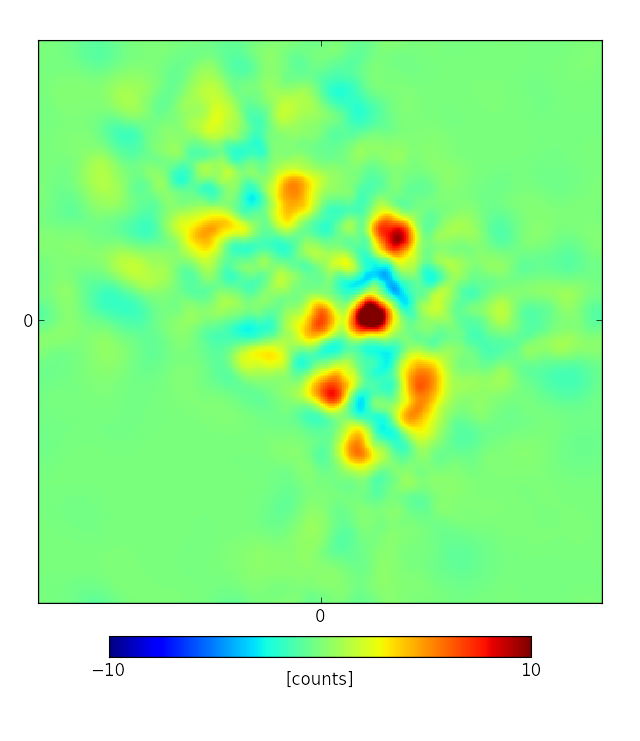}   \put(-3,97){(d)} \end{overpic} &
            \begin{overpic} [scale=0.35]{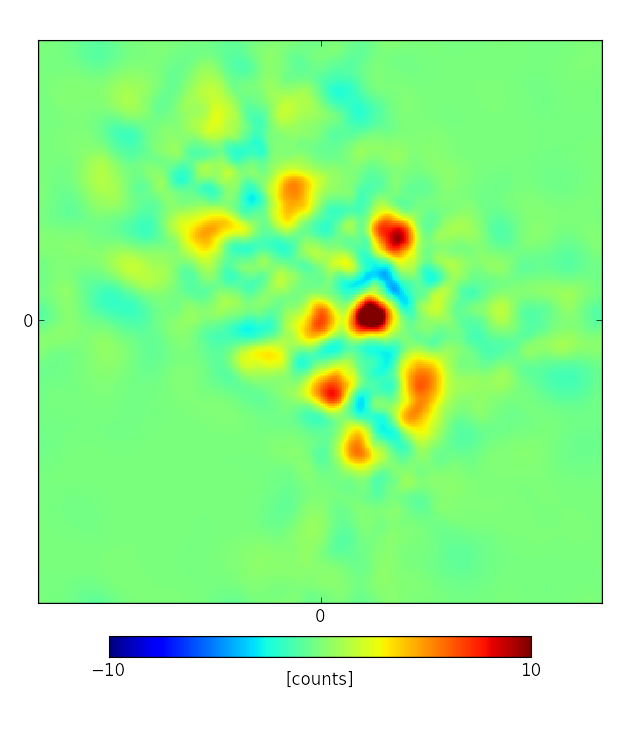}  \put(-3,97){(e)} \end{overpic} &
            \begin{overpic} [scale=0.35]{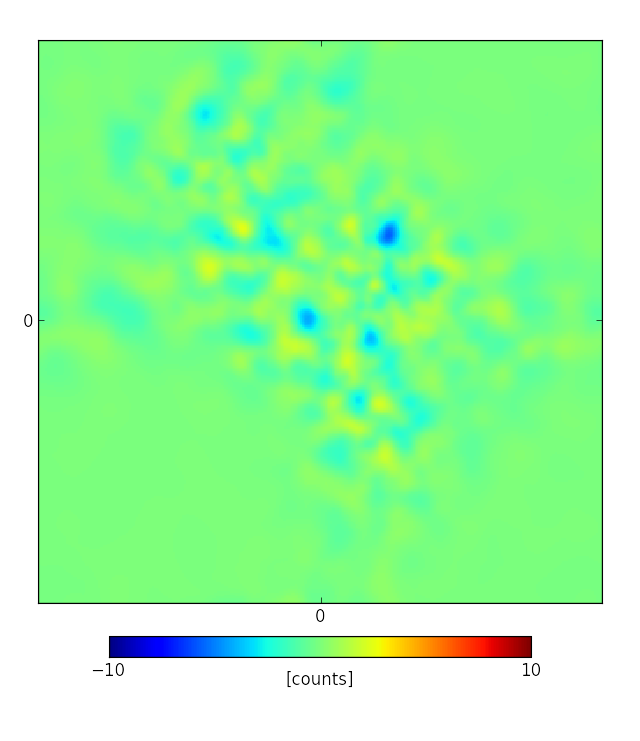} \put(-3,97){(f)} \end{overpic} \\
        \end{tabular}
        \flushleft
        \caption{Illustration of the diffuse reconstruction. The top panels show the denoised and deconvolved diffuse contribution $\langle\bb{\rho}^{(s)}\rangle/\rho_0$ reconstructed using a different approach each, namely (d) MAP-$\delta$, (e) MAP-$\G$, and (f) Gibbs. The bottom panels (d) to (f) show the difference between the originally simulated signal and the respective reconstruction.}
        \label{fig:rs}
    \end{figure*}

    \begin{figure*}[t]
        \centering
        \begin{tabular}{ccc}
            \begin{overpic} [scale=0.35]{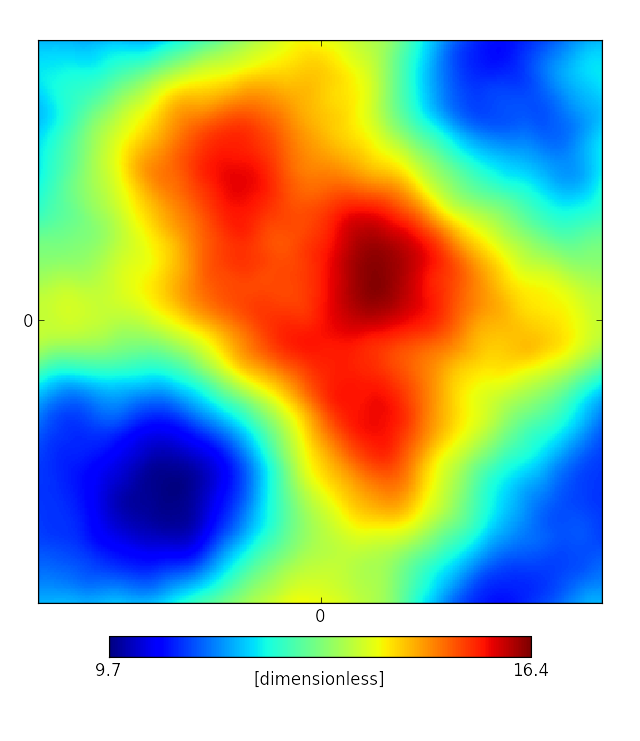}          \put(-3,97){(a)} \end{overpic} &
            \begin{overpic} [scale=0.35]{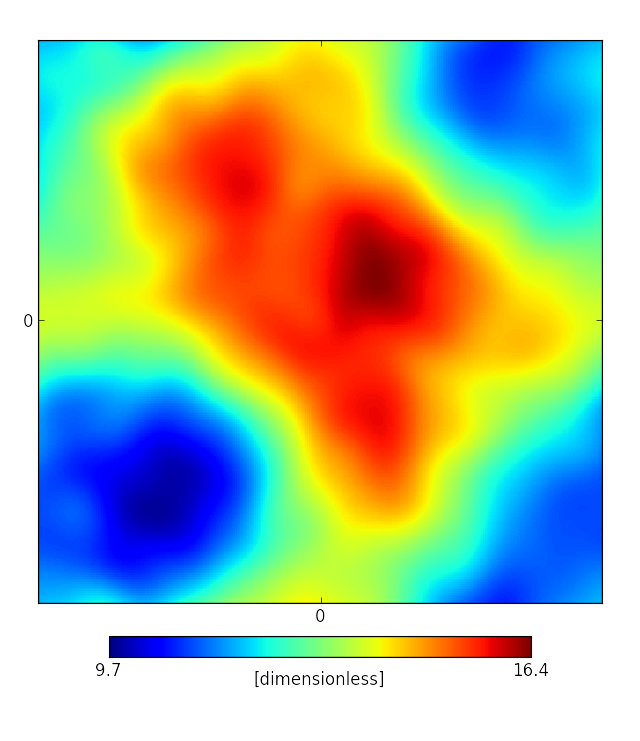}      \put(-3,97){(b)} \end{overpic} &
            \begin{overpic} [scale=0.35]{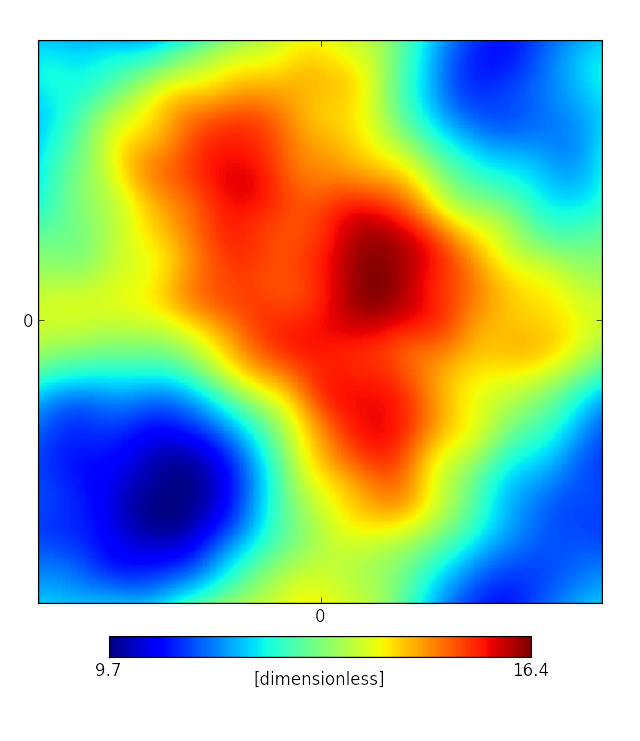}    \put(-3,97){(c)} \end{overpic} \\
            &
            \begin{overpic} [scale=0.35]{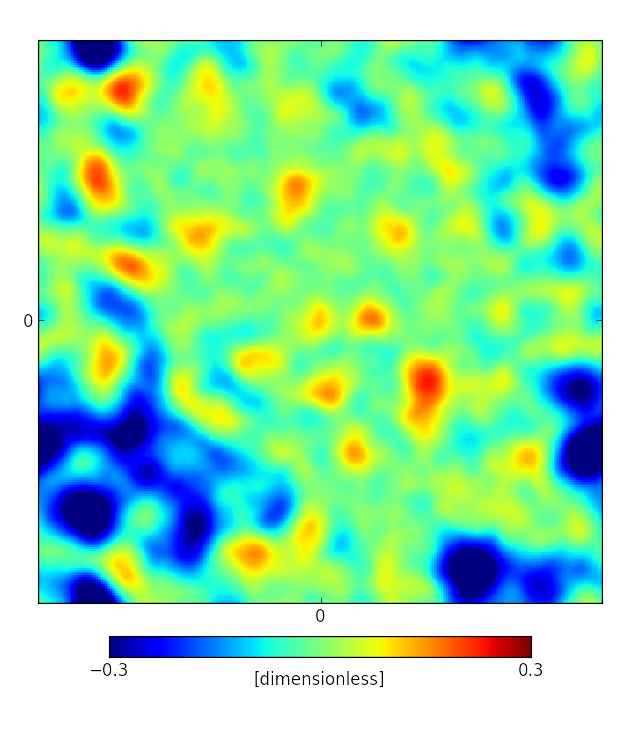}     \put(-3,97){(d)$\,=\,$(a)$\,-\,$(b)} \end{overpic} &
            \begin{overpic} [scale=0.35]{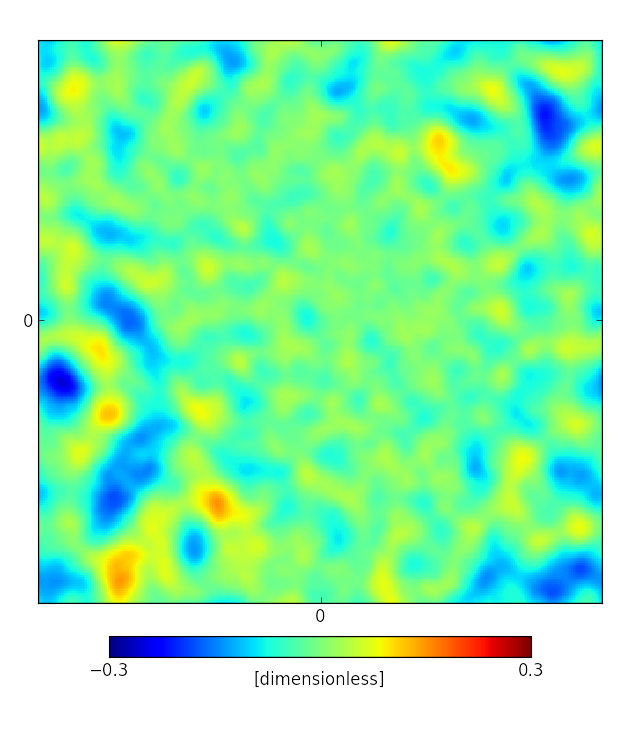}   \put(-3,97){(e)$\;=\,$(a)$\,-\,$(c)} \end{overpic} \\
            \begin{overpic} [scale=0.35]{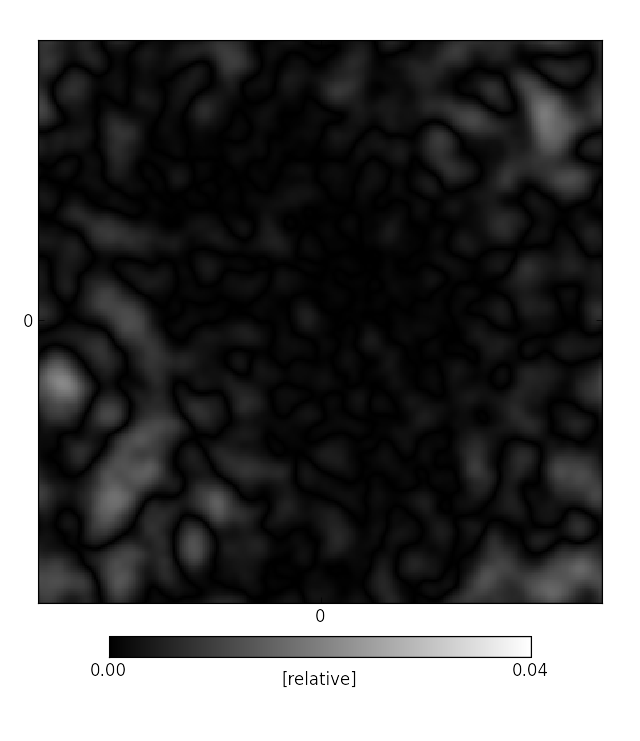} \put(-3,97){(f)$\;=|$(e)$|\,/\,$(c)} \end{overpic} &
            \begin{overpic} [scale=0.35]{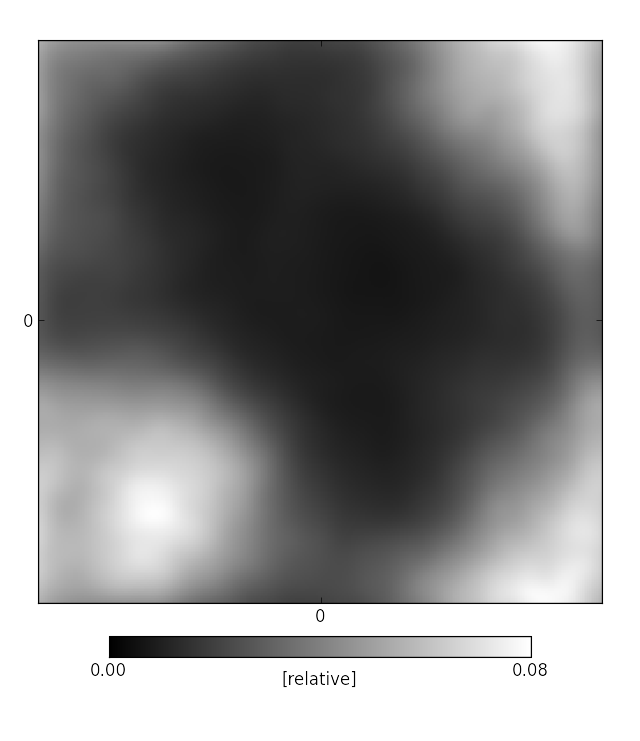}      \put(-3,97){(g)} \end{overpic} &
            \begin{overpic} [scale=0.35]{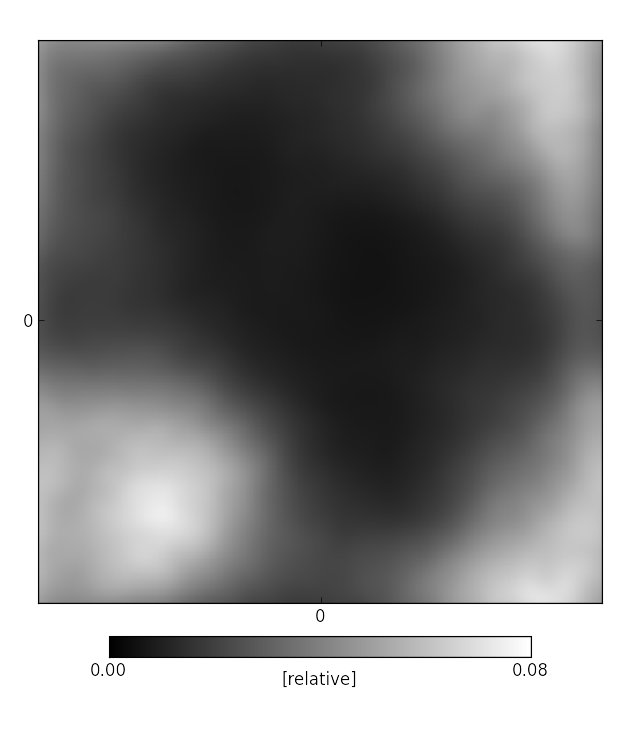}    \put(-3,97){(h)} \end{overpic} \\
        \end{tabular}
        \flushleft
        \caption{Illustration of the reconstruction of the diffuse signal field $\bb{s} = \log\bb{\rho}^{(s)}$ and its uncertainty. The top panels show diffuse signal fields. Panel (a) shows the original simulation $\bb{s}$, panel (b) the reconstruction $\bb{m}_\mathrm{mode}^{(s)}$ using a MAP approach, and panel (c) the reconstruction $\bb{m}_\mathrm{mean}^{(s)}$ using a Gibbs approach. The panels (d) and (e) show the differences between original and reconstruction. Panel (f) shows the relative difference. The panels (g) and (h) show the relative uncertainty of the above reconstructions.}
        \label{fig:s}
    \end{figure*}

    \begin{figure*}[!t]
        \centering
        \begin{tabular}{cc}
            \begin{overpic} [scale=0.5]{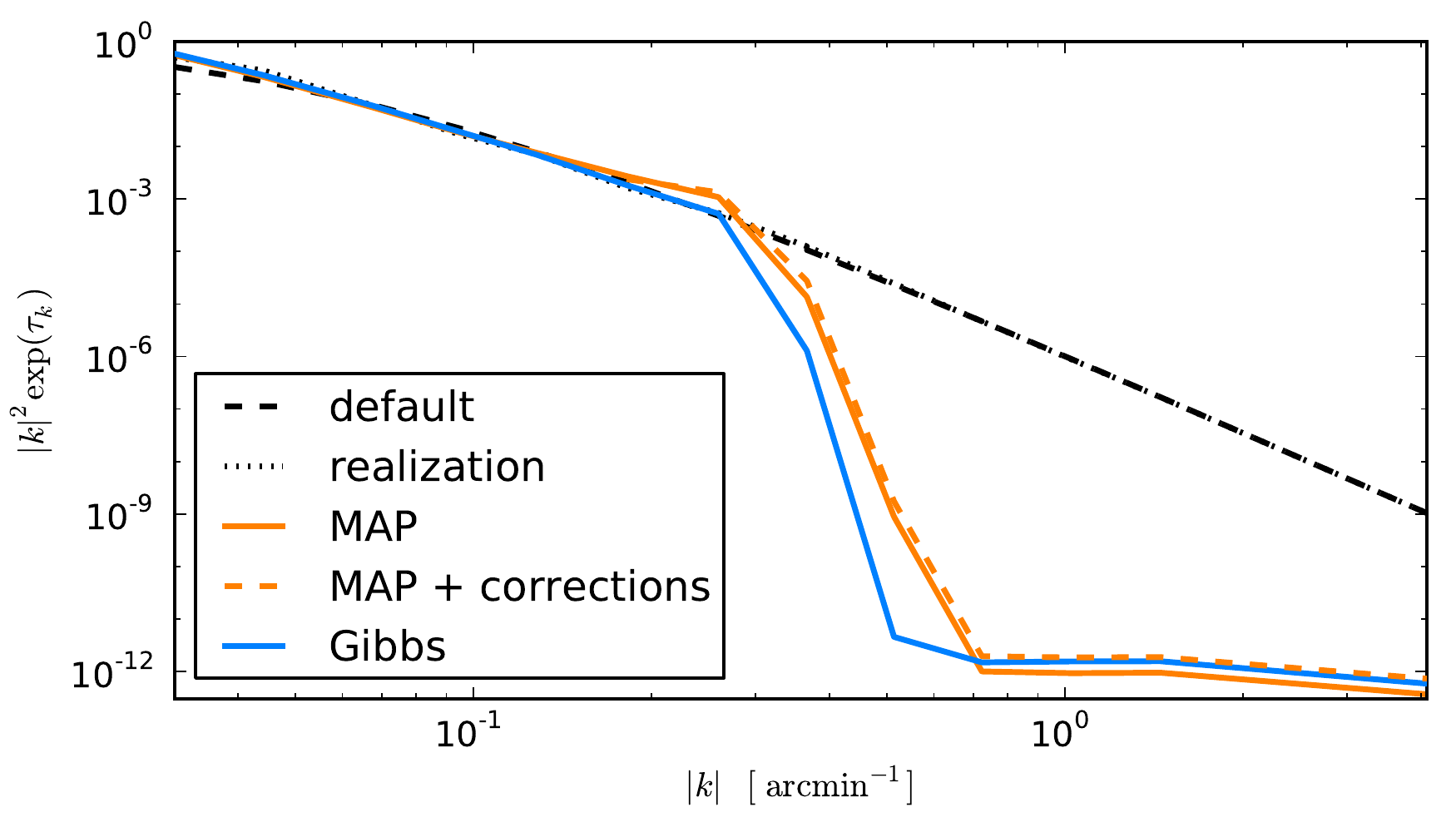}   \put(-3,57){(a)} \end{overpic} &
            \begin{overpic} [scale=0.5]{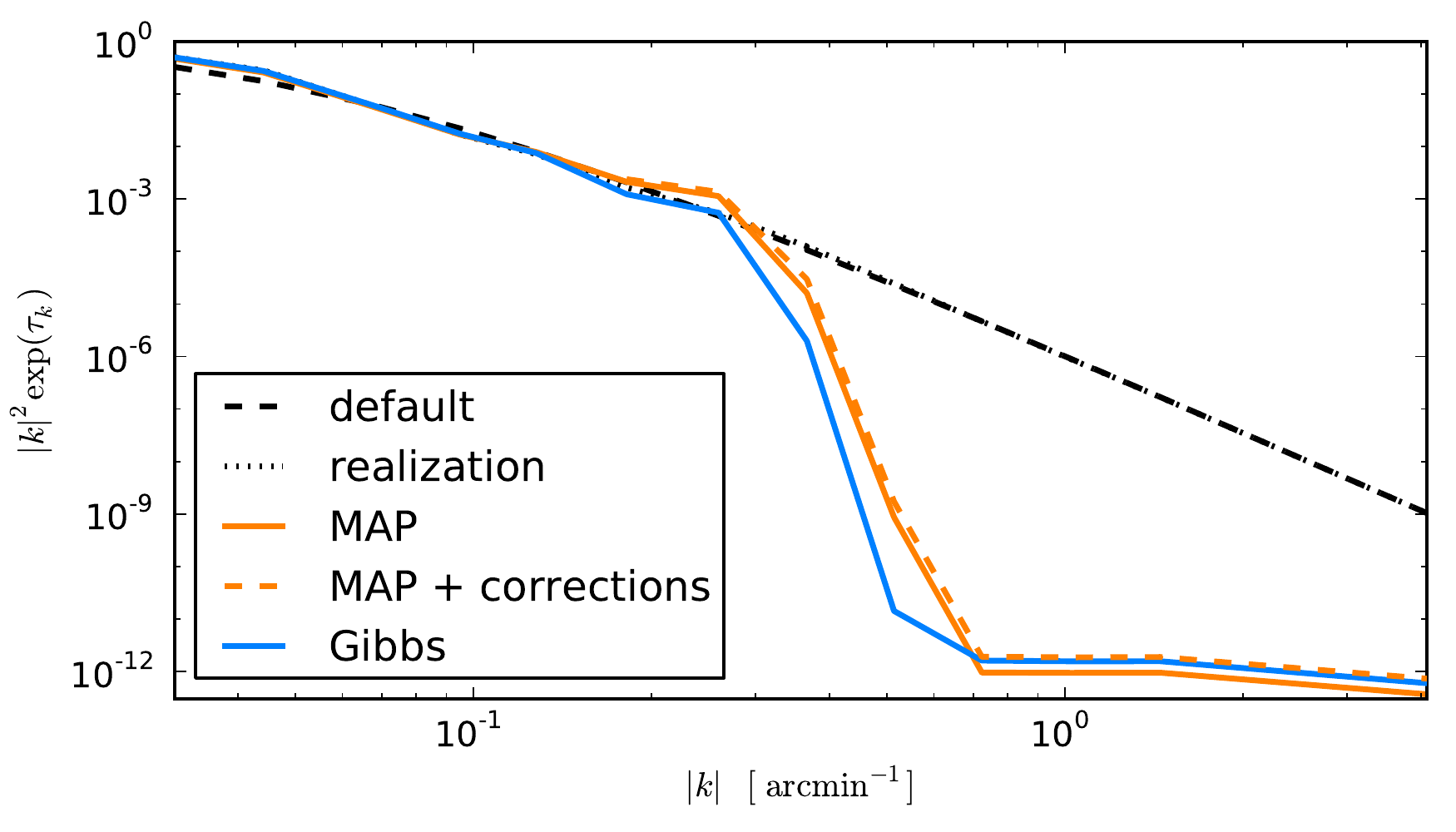} \put(-3,57){(b)} \end{overpic} \\
        \end{tabular}
        \flushleft
        \caption{Illustration of the reconstruction of the logarithmic power spectrum $\bb{\tau}$. Both panels show the default power spectrum (black dashed line), and the simulated realization (black dotted line), as well as the reconstructed power spectra using a MAP (orange solid line), plus second order corrections (orange dashed line), and a Gibbs approach (blue solid line). Panel (a) shows the reconstruction for a chosen $\sigma$ parameter of $10$, panel (b) for a $\sigma$ of $1000$.}
        \label{fig:t}
    \end{figure*}

    \begin{figure*}[t]
        \centering
        \begin{tabular}{ccc}
            \begin{overpic} [scale=0.35]{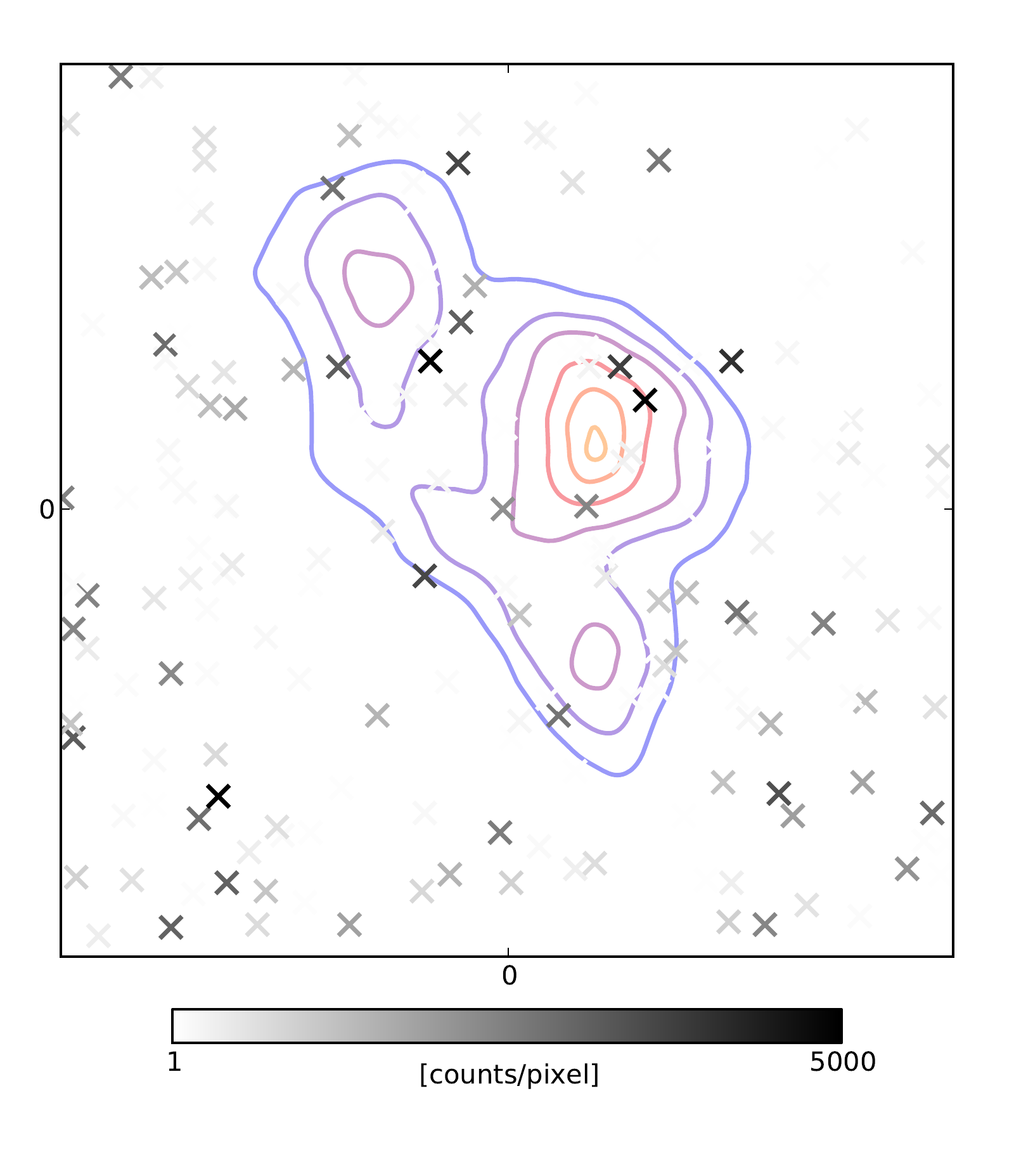}       \put(-3,97){(a)} \end{overpic} &
            \begin{overpic} [scale=0.35]{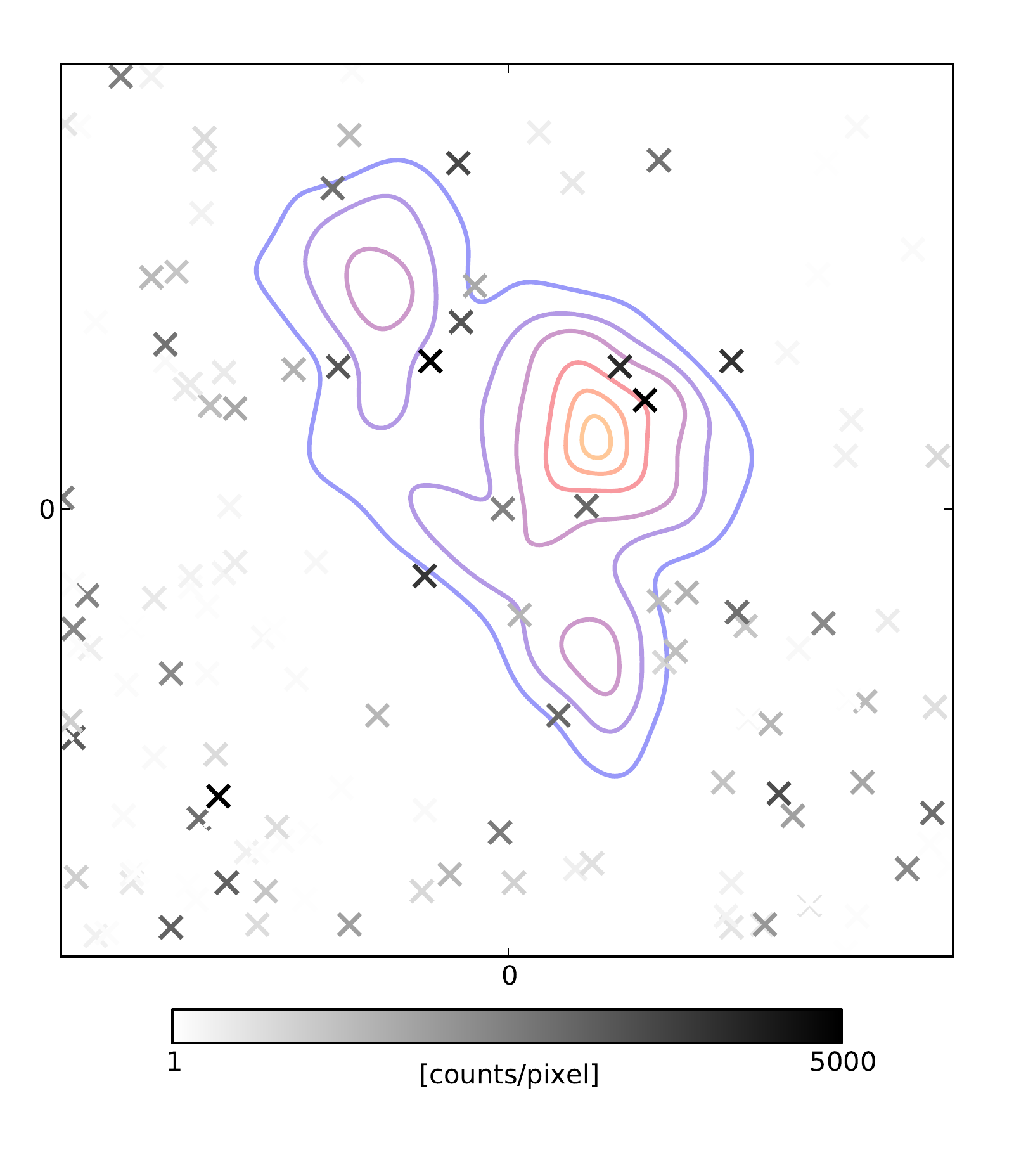}   \put(-3,97){(b)} \end{overpic} &
            \begin{overpic} [scale=0.35]{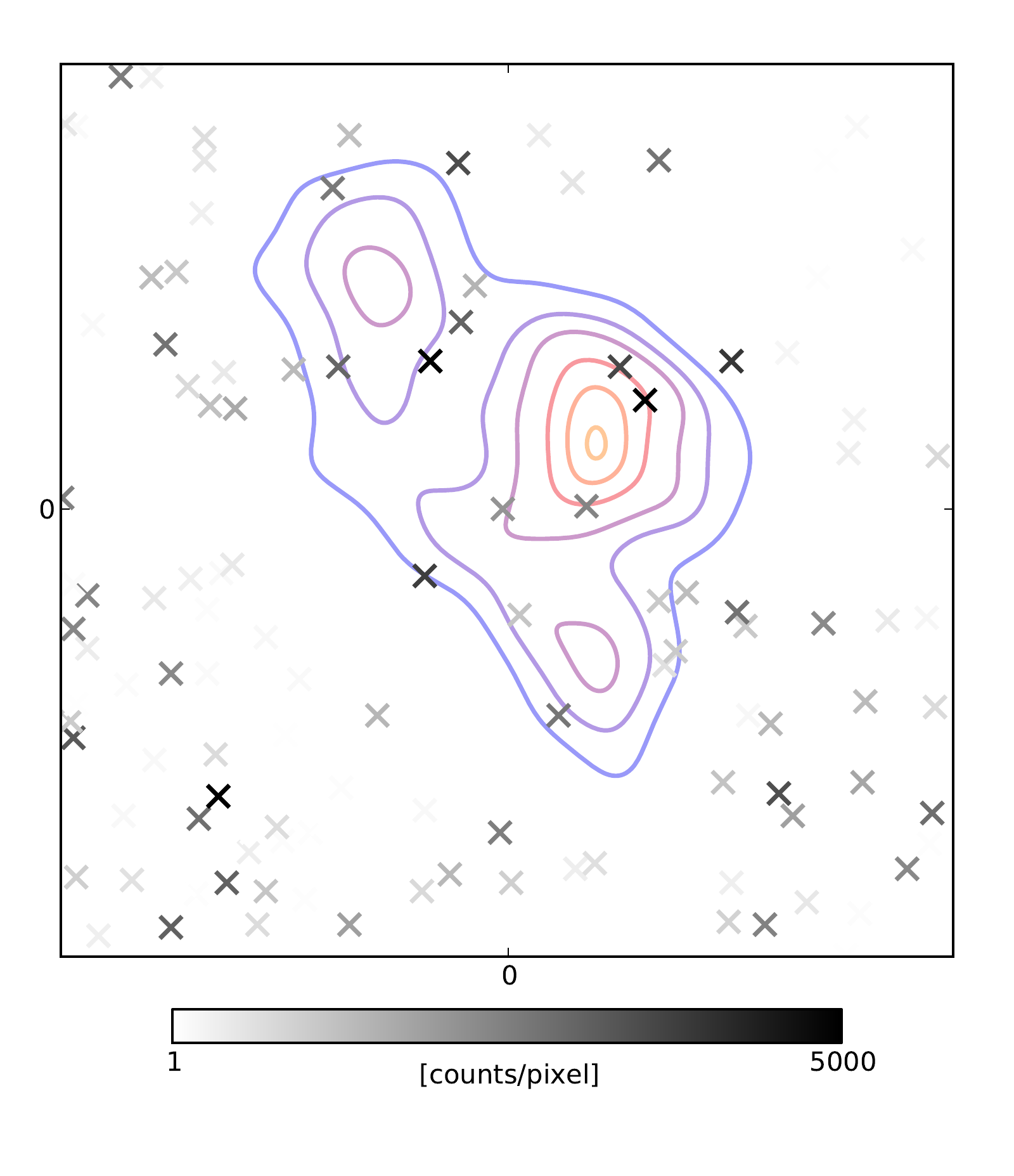} \put(-3,97){(c)} \end{overpic} \\
            &
            \begin{overpic} [scale=0.35]{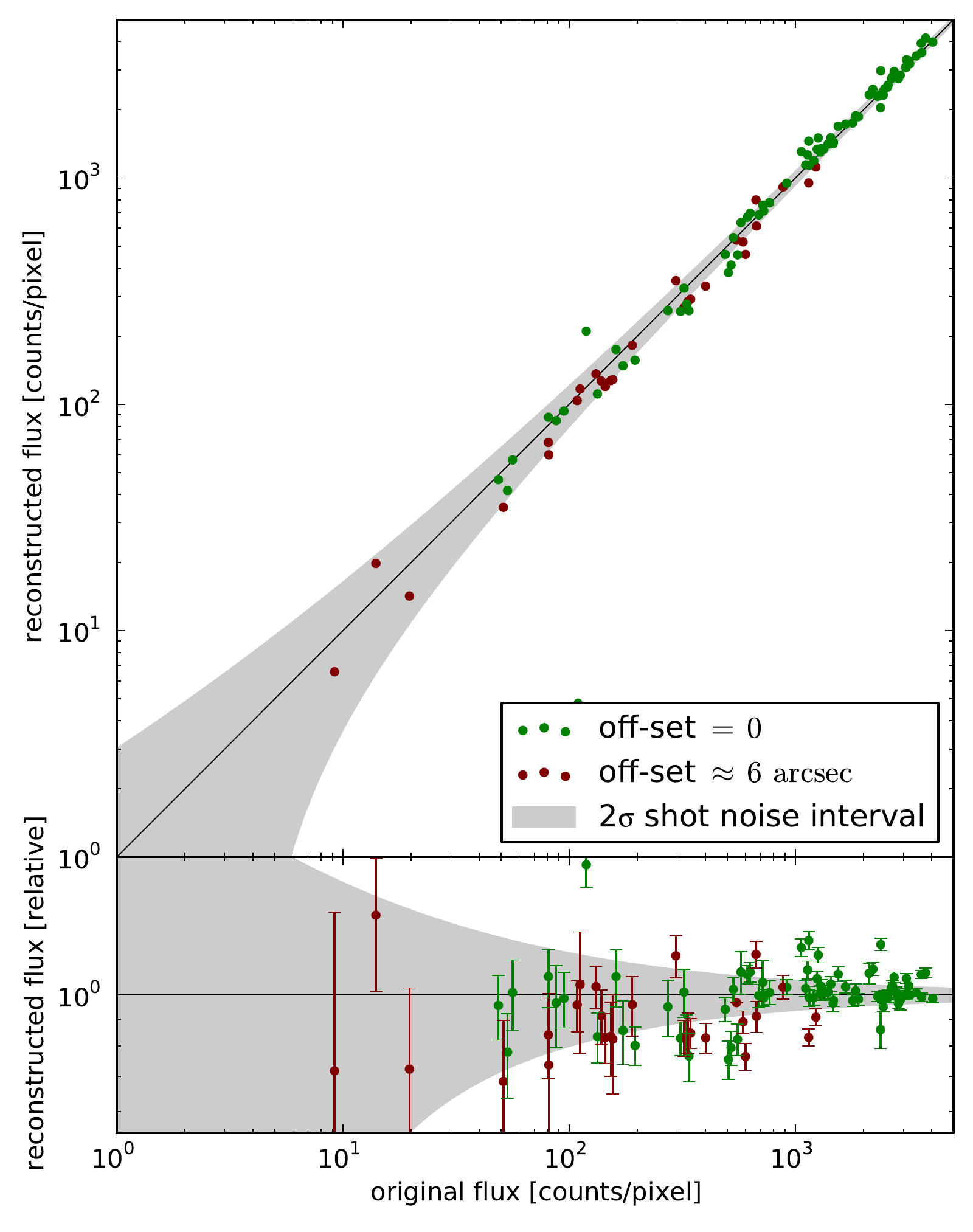}      \put(-3,100){(d)} \end{overpic} &
            \begin{overpic} [scale=0.35]{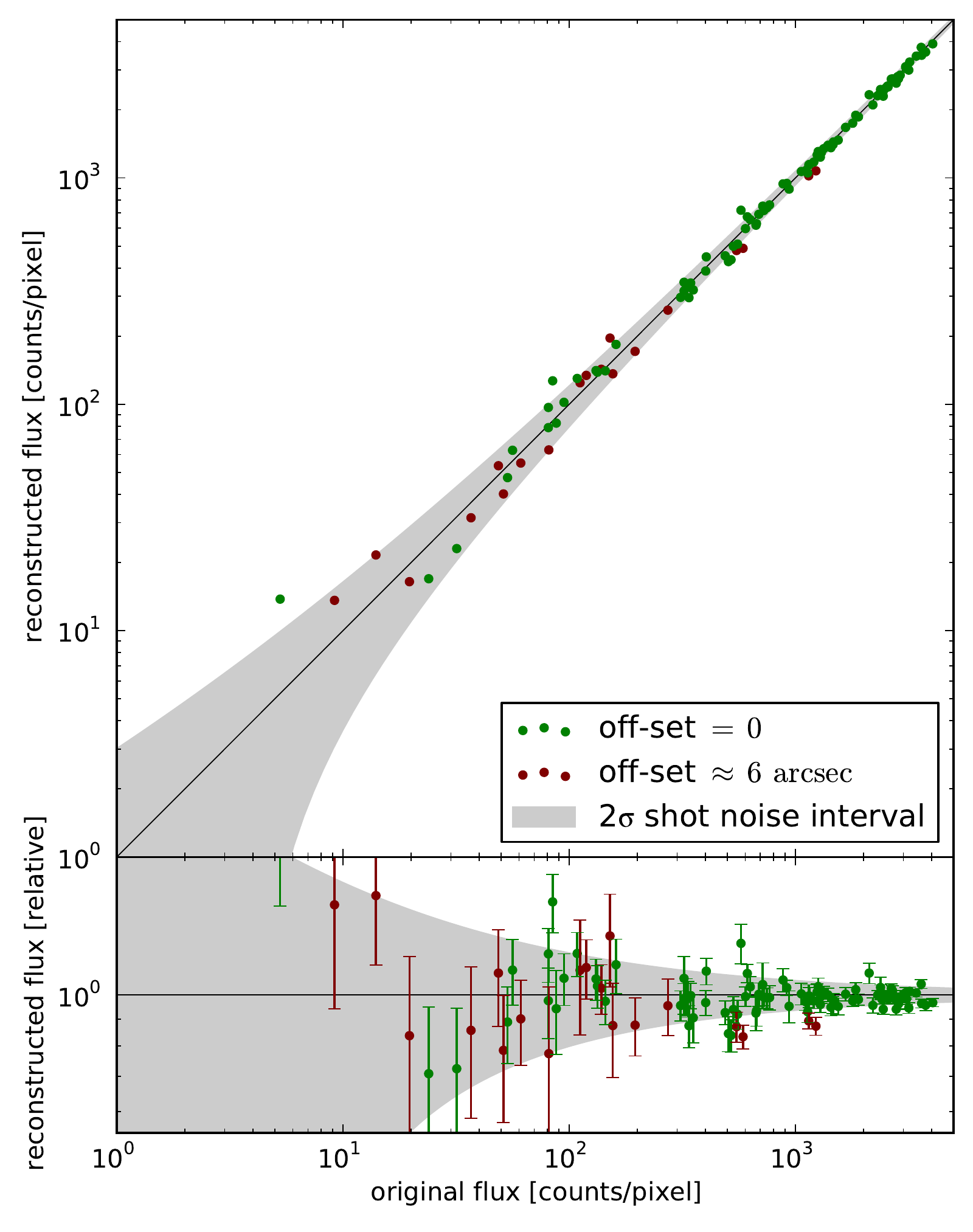}    \put(-3,100){(e)} \end{overpic} \\
        \end{tabular}
        \flushleft
        \caption{Illustration of the reconstruction of the point-like signal field $\bb{u} = \log\bb{\rho}^{(u)}$ and its uncertainty. The top panels show the location (markers) and intensity (gray scale) of the point-like photon fluxes, underlaid is the respective diffuse contribution (contours) to guide the eye, cf. Fig~\ref{fig:rs}. Panel (a) shows the original simulation, panel (b) the reconstruction using a MAP approach, and panel (c) the reconstruction using a Gibbs approach. The bottom panels (d) and (e) show the match between original and reconstruction in absolute and relative fluxes, the $2\sigma$ shot noise interval (gray contour), as well as some reconstruction uncertainty estimate (error bars).}
        \label{fig:u}
    \end{figure*}

    The problem of denoising, deconvolving, and decomposing photon observations is a non-trivial task. Therefore, this section discusses the implementation of the D$^3$PO algorithm given the two sets of filter formulas derived in Sec.~\ref{sec:MAP} and \ref{sec:Gibbs}, respectively.

    The information Hamiltonian, or equivalently the Gibbs free energy, are scalar quantities defined over a huge phase space of possible field and parameter configurations including, among others, the elements of $\bb{m}^{(s)}$ and $\bb{m}^{(u)}$. If we only consider those, and no resolution refinement from data to signal space, two numbers need to be inferred from one data value each. Including $\bb{\tau}$ and the uncertainty covariances $\bb{D}^{(s)}$ and $\bb{D}^{(u)}$ in the inference, the problem of underdetermined degrees of freedom gets worse. This is reflected in the possibility of a decent number of local minima in the non-convex manifold landscape of the codomain of the Hamiltonian, or Gibbs free energy, respectively \citep{KGV83,GG84,GC08}. The complexity of the inference problem goes back to the, in general, non-linear entanglement between the individual parameters.

    The D$^3$PO algorithm is based on an iterative optimization scheme, where certain subsets of the problem are optimized alternately instead of the full problem at once. Each subset optimization is designed individually, see below.
    The global optimization cycle is in some degree sensitive to the starting values because of the non-convexity of the considered potential; i.e., the information Hamiltonian or Gibbs free energy, respectively. We can find such appropriate starting values by solving the inference problem in a reduced frame in advance, see below.
    So far, a step-by-step guide of the algorithm looks like the following.

    \vspace{0.5em}
    \begin{enumerate}%[label={(\arabic*)}]

        \item \label{i}
            Initialize the algorithm with primitive starting values; e.g., $m_x^{(s)} = m_x^{(u)} = 0$, $D_{xy}^{(s)} = D_{xy}^{(u)} = \delta_{xy}$, and $\tau_k^\star = \log(k^{-2})$. -- Those values are arbitrary. Although the optimization is rather insensitive to them, inappropriate values can cripple the algorithm for numerical reasons because of the high non-linearity of the inference problem.

        \item \label{s0}
            Optimize $\bb{m}^{(s)}$, the diffuse signal field, coarsely. -- The preliminary optimization shall yield a rough estimate of the diffuse only contribution. This can be achieved by reconstructing a coarse screened diffuse signal field that only varies on large scales; i.e., limiting the bandwidth of the diffuse signal in its harmonic basis. Alternatively, obvious point sources in the data could be masked out by introducing an artificial mask into the response, if feasible.

        \item \label{u0}
            Optimize $\bb{m}^{(u)}$, the point-like signal field, locally. -- This initial optimization shall approximate the brightest, most obvious, point sources that are visible in the data image by eye. Their current disagreement with the data dominates the considered potential, and introduces some numerical stiffness.
            The gradient of the potential can be computed according to Eq.~\eqref{eq:u_MAP} or \eqref{eq:u_Gibbs}, and its minima will be at the expected position of the brightest point source which has not been reconstructed, yet. It is therefore very efficient to increase $\bb{m}^{(u)}$ at this location directly until the sign of the gradient flips, and repeat this procedure until the obvious point sources are fit.

        \item \label{u1}
            Optimize $\bb{m}^{(u)}$, the point-like signal field. --
            This task can be done by a steepest descent minimization of the potential combined with a line search following the Wolfe conditions \citep{NW06}. The potentials can be computed according to Eq.~\eqref{eq:H} or \eqref{eq:G} neglecting terms independent of $\bb{m}^{(u)}$, and the gradient according to Eq.~\eqref{eq:u_MAP} or \eqref{eq:u_Gibbs}. A more sophisticated minimization scheme, such as a non-linear conjugate gradient \citep{S94}, is conceivable but would require the application of the full Hessian, cf. step~\ref{u2}.
            In the first run, it might be sufficient to restrict the optimization to the locations identified in step~\ref{u0}.

        \item \label{u2}
            Update $\hh{D}^{(u)}$, the point-like uncertainty variance, in case of a Gibbs approach. --
            It is not feasible to compute the full uncertainty covariance $\bb{D}^{(u)}$ explicitly in order to extract its diagonal.
            A more elegant way is to apply a probing technique relying on the application of $\bb{D}^{(u)}$ to random fields $\bb{\xi}$ that project out the diagonal \citep{H89,SOE12}. The uncertainty covariance is given as the inverse Hessian by Eq.~\eqref{eq:D_MAP} or \eqref{eq:D_Gibbs}, and should be symmetric and positive definite. For that reason, it can be applied to a field using a conjugate gradient \citep{S94}; i.e., solving $(\bb{D}^{(u)})^{-1}\bb{y} = \bb{\xi}$ for $\bb{y}$.
            However, if the current phase space position is far away from the minimum, the Hessian is not necessarily positive definite. One way to overcome this temporal instability, would be to introduce a Levenberg damping in the Hessian \citep[inspired by][]{TMS09,TS12}.

        \item \label{s1}
            Optimize $\bb{m}^{(s)}$, the diffuse signal field. --
            An analog scheme as in step~\ref{u1} using steepest descent and Wolfe conditions is effective.
            The potentials can be computed according to Eq.~\eqref{eq:H} or \eqref{eq:G} neglecting terms independent of $\bb{m}^{(s)}$, and the gradient according to Eq.~\eqref{eq:s_MAP} or \eqref{eq:s_Gibbs}, respectively. It has proven useful to first ensure a convergence on large scales; i.e., small harmonic modes $k$. This can be done repeating steps~\ref{s1}, \ref{s2}, and \ref{t} for all $k < k_\mathrm{max}$ with growing $k_\mathrm{max}$ using the corresponding projections $\bb{S}_k$.

        \item \label{s2}
            Update $\hh{D}^{(s)}$, the diffuse uncertainty variance, in case of a Gibbs approach in analogy to step~\ref{u2}.

        \item \label{t}
            Optimize $\bb{\tau}^\star$, the logarithmic power spectrum. --
            This is done by solving Eq.~\eqref{eq:t_MAP} or \eqref{eq:t_Gibbs}. The trace term can be computed analog to the diagonal; e.g., by probing. Given this, the equation can be solved efficiently by a Newton-Raphson method.

        \item
            Repeat the steps~\ref{u1} to \ref{t} until convergence. --
            This scheme will take several cycles until the algorithm reaches the desired convergence level. Therefore, it is not required to achieve a convergence to the final accuracy level in all subsets in all cycles. It is advisable to start with weak convergence criteria in the first loop and increase them gradually.

    \end{enumerate}
    %\vspace{0.5em}

    \noindent
    A few remarks are in order.

    The phase space of possible signal field configurations is tremendously huge. It is therefore impossible to judge if the algorithm has converged to the global or some local minima, but this does not matter if both yield reasonable results that do not differ substantially.

    In general, the converged solution is also subject to the choice of starting values. Solving a non-convex, non-linear inference problem without proper initialization can easily lead to nonsensical results, such as fitting (all) diffuse features by point sources.
    Therefore, the D$^3$PO algorithm essentially creates its own starting values executing the initial steps~\ref{i} to \ref{u0}. The primitive starting values are thereby processed to rough estimates that cover coarsely resolved diffuse and prominent point-like features. These estimates serve then as actual starting values for the optimization cycle.

    Because of the iterative optimization scheme starting with the diffuse component in step~\ref{s0}, the algorithm might be prone to explaining some point-like features by diffuse sources. Starting with the point-like component instead would give rise to the opposite bias.
    To avoid such biases, it is advisable to restart the algorithm partially. To be more precise, we propose to discard the current reconstruction of $\bb{m}^{(u)}$ after finishing step~\ref{t} for the first time, then start the second iteration again with step~\ref{u0}, and to discard the current $\bb{m}^{(s)}$ before step~\ref{s1}.

    The above scheme exploits a few numerical techniques, such as probing or Levenberg damping, that are described in great detail in the given references. The code of our implementation of the D$^3$PO algorithm will be made public in the future under \url{http://www.mpa-garching.mpg.de/ift/d3po/}.

%%================================
\section{Numerical application}
\label{sec:application}

    Exceeding the simple 1D scenario illustrated in Fig.~\ref{fig:motivation}, the D$^3$PO algorithm is now applied to a realistic, but simulated, data set. The data set represents a high energy observation with a field of view of $32 \times 32 \,\mathrm{arcmin}^2$ and a resolution of $0.1 \,\mathrm{arcmin}$; i.e., the photon count image comprises $102\,400$ pixels. The instrument response includes the convolution with a Gaussian-like PSF with a FWHM of roughly $0.2 \,\mathrm{arcmin}$, and an uneven survey mask attributable to the inhomogeneous exposure of the virtual instrument. The data image and those characteristics are shown in Fig.~\ref{fig:d}.

    In addition, the top panels of Fig.~\ref{fig:d} show the reproduced signal responses of the reconstructed (total) photon flux. The reconstructions used the same model parameters, $\alpha=1$, $q=10^{-12}$, $\sigma=10$, $\beta = \tfrac{3}{2}$, and $\eta = 10^{-4}$ in a MAP-$\delta$, MAP-$\G$ and a Gibbs approach, respectively. They all show a very good agreement with the actual data, and differences are barely visible by eye.
    We note that only the quality of denoising is visible, since the signal response shows the convolved and superimposed signal fields.

    The diffuse contribution to the deconvolved photon flux is shown Fig.~\ref{fig:rs} for all three estimators, cf. Eqs.~\eqref{eq:r_MAP} to \eqref{eq:r_Gibbs}. There, all point-like contributions as well as noise and instrumental effects have been removed presenting a denoised, deconvolved and decomposed reconstruction result for the diffuse photon flux.
    Figure~\ref{fig:rs} also shows the absolute difference to the original flux. Although the differences in the MAP estimators are insignificant, the Gibbs solution seems to be slightly better.

    In order to have a quantitative statement about the goodness of the reconstruction, we define a relative residual error $\epsilon^{(s)}$ for the diffuse contribution as follows,
    \begin{align}
        \epsilon^{(s)} &= \left| \bb{\rho}^{(s)} - \left< \bb{\rho}^{(s)} \right> \right|_2 \left| \bb{\rho}^{(s)} \right|_2^{-1}
        , \label{eq:errs}
    \end{align}
    where $|\,\cdot\,|_2$ is the Euclidean L$^2$-norm. For the point-like contribution, however, we have to consider an error in brightness and position. For this purpose we define,
    \begin{align}
        \epsilon^{(u)} &= \int_1^N \d n \; \left| \bb{R}_\mathrm{PSF}^n \bb{\rho}^{(u)} - \bb{R}_\mathrm{PSF}^n \left< \bb{\rho}^{(u)} \right> \right|_2 \left| \bb{R}_\mathrm{PSF}^n \bb{\rho}^{(u)} \right|_2^{-1}
        , \label{eq:erru}
    \end{align}
    where $\bb{R}_\mathrm{PSF}$ is a (normalized) convolution operator, such that $\bb{R}_\mathrm{PSF}^N$ becomes the identity for large $N$.
    These errors are listed in Table~\ref{tab:errs}. When comparing the MAP-$\delta$ and MAP-$\G$ approach, the incorporation of uncertainty corrections seems to improve the results slightly. The full regularization treatment within the Gibbs approach outperforms MAP solutions in terms of the chosen error measure $\epsilon^{(\,\cdot\,)}$.
    For a discussion of how such measures can change the view on certain Bayesian estimators, we refer to the work by \citet{BL14}.

    \begin{table}[!b]
        \caption{Overview of the relative residual errors in the photon flux reconstructions for the respective approaches, all using the same model parameters, cf. text.}
        \centering
        \begin{tabular}{|rrr|}
            \hline
            \multicolumn{1}{|c}{MAP-$\delta$} & \multicolumn{1}{c}{MAP-$\G$} & \multicolumn{1}{c|}{Gibbs} \\
            \hline
            \hline
            $\epsilon^{(s)} = 4.442$\% & $\epsilon^{(s)} = 4.441$\% & $\epsilon^{(s)} = 2.078$\% \\%   4.44190179   4.44112794   2.07831735
            $\epsilon^{(u)} = 1.540$\% & $\epsilon^{(u)} = 1.540$\% & $\epsilon^{(u)} = 1.089$\% \\%   1.53970515   1.53970516   1.08854039
            \hline
        \end{tabular}
        \label{tab:errs}
    \end{table}

    Figure~\ref{fig:s} illustrates the reconstruction of the diffuse signal field, now in terms of logarithmic flux.
    The original and the reconstructions agree well, and the strongest deviations are found in the areas with low amplitudes. With regard to the exponential ansatz in Eq.~\eqref{eq:superposition}, it is not surprising that the inference on the signal fields is more sensitive to higher values than to lower ones. For example, a small change in the diffuse signal field, $\bb{s} \rightarrow (1 \pm \epsilon) \bb{s}$, translates into a factor in the photon flux, $\bb{\rho}^{(s)} \rightarrow \bb{\rho}^{(s)} \e^{\pm\epsilon\bb{s}}$, that scales exponentially with the amplitude of the diffuse signal field.
    The Gibbs solution shows less deviation from the original signal than the MAP solution. Since the latter lacks the regularization by the uncertainty covariance it exhibits a stronger tendency to overfitting compared to the former. This includes overestimates in noisy regions with low flux intensities, as well as underestimates at locations where point-like contributions dominate the total flux.%The latter indicates an overfitting of the point-like component by the MAP approach.

    The reconstruction of the power spectrum, as shown in Fig.~\ref{fig:t}, gives further indications of the reconstruction quality of the diffuse component. The simulation used a default power spectrum of
    \begin{align}
        \exp(\tau_k)&= 42 \, (k+1)^{-7}
        .
    \end{align}
    This power spectrum was on purpose chosen to deviate from a strict power law supposed by the smoothness prior.

    From Fig.~\ref{fig:t} it is apparent that the reconstructed power spectra track the original well up to a harmonic mode $k$ of roughly $0.4 \,\mathrm{arcmin}^{-1}$. Beyond that point, the reconstructed power spectra fall steeply until they hit a lower boundary set by the model parameter $q = 10^{-12}$. This drop-off point at $0.4 \,\mathrm{arcmin}^{-1}$ corresponds to a physical wavelength of roughly $2.5 \,\mathrm{arcmin}$, and thus (half-phase) fluctuations on a spatial distances below $1.25 \,\mathrm{arcmin}$. The Gaussian-like PSF of the virtual observatory has a finite support of $1.1 \,\mathrm{arcmin}$. The lack of reconstructed power indicates that the algorithm assigns features on spatial scales smaller than the PSF support preferably to the point-like component.
    This behavior is reasonable because solely the point-like signal can cause PSF-like shaped imprints in the data image. However, there is no strict threshold in the distinction between the components on the mere basis of their spatial extend. We rather observe a continuous transition from assigning flux to the diffuse component to assigning it to the point-like component while reaching smaller spatial scales because strict boundaries are blurred out under the consideration of noise effects.

    The differences between the reconstruction using a MAP and a Gibbs approach are subtle.
    The difference in the reconstruction formulas given by Eqs.~\eqref{eq:t_MAP} and \eqref{eq:t_Gibbs} is an additive trace term involving $\bb{D}^{(s)}$, which is positive definite. Therefore, a reconstructed power spectrum regularized by uncertainty corrections is never below the one with out given the same $\bb{m}^{(s)}$.
    However, the reconstruction of the signal field follows different filter formulas, respectively. Since the Gibbs approach considers the uncertainty covariance $\bb{D}^{(s)}$ properly in each cycle, it can present a more conservative solution.
    The drop-off point is apparently at higher $k$ for the MAP approach, leading to higher power on scales between roughly $0.3$ and $0.7 \,\mathrm{arcmin}^{-1}$.
    In turn, the MAP solution tends to overfit by absorbing some noise power into $\bb{m}^{(s)}$ as discussed in Sec.~\ref{sec:solution}. Thus, the higher MAP power spectrum in Fig.~\ref{fig:t} seems to be caused by a higher level of noise remnants in the signal estimate.

    The influence of the choice of the model parameter $\sigma$ is also shown in Fig.~\ref{fig:t}. Neither a smoothness prior with $\sigma = 10$, nor a weak one with $\sigma = 1000$ influences the reconstruction of the power spectrum substantially in this case.\footnote{For a discussion of further log-normal reconstruction scenarios please refer to the work by \citet{OSBE12}.} The latter choice, however, exhibits some more fluctuations in order to better track the concrete realization.

    The results for the reconstruction of the point-like component are illustrated in Fig.~\ref{fig:u}. Overall, the reconstructed point-like signal field and the corresponding photon flux are in good agreement with the original ones. The point-sources have been located with an accuracy of $\pm 0.1 \,\mathrm{arcmin}$, which is less than the FWHM of the PSF. The localization tends to be more precise for higher flux values because of the higher signal-to-noise ratio. The reconstructed intensities match the simulated ones well, although the MAP solution shows a spread that exceeds the expected shot noise uncertainty interval. This is again an indication of the overfitting known for MAP solutions. Moreover, neither reconstruction shows a bias towards higher or lower fluxes.

    The uncertainty estimates for the point-like photon flux $\bb{\rho}^{(u)}$ obtained from $\bb{D}^{(u)}$ according to Eqs.~\eqref{eq:var_MAP} and \eqref{eq:var_Gibbs} are, in general, consistent with the deviations from the original and the shot noise uncertainty, cf. Fig.~\ref{fig:u}. They show a reasonable scaling being higher for lower fluxes and vice versa.
    However, some uncertainties seem to be underestimated. There are different reasons for this.

    On the one hand, the Hessian approximation for $\bb{D}^{(u)}$ in Eq.~\eqref{eq:D_MAP} or \eqref{eq:D_Gibbs} is in individual cases in so far poor as that the curvature of the considered potential does not describe the uncertainty of the point-like component adequately. The data admittedly constrains the flux intensity of a point source sufficiently, especially if it is a bright one. However, the rather narrow dip in the manifold landscape of the considered potential can be asymmetric, and thus not always well described by the quadratic approximation of Eq.~\eqref{eq:D_MAP} or \eqref{eq:D_Gibbs}, respectively.

    On the other hand, the approximation leading to vanishing cross-correlation $\bb{D}^{(su)}$, takes away the possibility of communicating uncertainties between diffuse and point-like components.
    However, omitting the used simplification or incorporating higher order corrections would render the algorithm too computationally expensive. The fact that the Gibbs solution, which takes $\bb{D}^{(u)}$ into account, shows improvements backs up this argument.

    The reconstructions shown in Fig.~\ref{fig:s} and \ref{fig:u} used the model parameters $\sigma=10$, $\beta = \tfrac{3}{2}$, and $\eta = 10^{-4}$. In order to reflect the influence of the choice of $\sigma$, $\beta$, and $\eta$, Table~\ref{tab:runs} summarizes the results from several reconstructions carried out with varying model parameters. Accordingly, the best parameters seem to be $\sigma = 10$, $\beta = \tfrac{5}{4}$, and $\eta = 10^{-4}$, although we caution that the total error is difficile to determine as the residual errors, $\epsilon^{(s)}$ and $\epsilon^{(u)}$, are defined differently. Although the errors vary significantly, $2$--$15\%$ for $\epsilon^{(s)}$, we like to stress that the model parameters were changed drastically, partly even by orders of magnitude. The impact of the prior clearly exists, but is moderate.
    We note that the case of $\sigma \rightarrow \infty$ corresponds to neglecting the smoothness prior completely. The $\beta = 1$ case that corresponds to a logarithmically flat prior on $\bb{u}$ showed a tendency to fit more noise features by point-like contributions.

    In summary, the D$^3$PO algorithm is capable of denoising, deconvolving and decomposing photon observations by reconstructing the diffuse and point-like signal field, and the logarithmic power spectrum of the former. The reconstruction using MAP and Gibbs approaches perform flawlessly, except for a little underestimation of the uncertainty of the point-like component. The MAP approach shows signs of overfitting, but those are not overwhelming. Considering the simplicity of the MAP approach that goes along with a numerically faster performance, this shortcoming seems acceptable.

    Because of the iterative scheme of the algorithm, a combination of the MAP approach for the signal fields and a Gibbs approach for the power spectrum is possible.

%%================================
\section{Conclusions \& summary}
\label{sec:conclusion}

    The D$^3$PO algorithm for the denoising, deconvolving and decomposing photon observations has been derived. It allows for the simultaneous but individual reconstruction of the diffuse and point-like photon fluxes, as well as the harmonic power spectrum of the diffuse component, from a single data image that is exposed to Poissonian shot noise and effects of the instrument response functions. Moreover, the D$^3$PO algorithm can provide \emph{a~posteriori} uncertainty information on the reconstructed signal fields. With these capabilities, D$^3$PO surpasses previous approaches that address only subsets of these complications.

    The theoretical foundation is a hierarchical Bayesian parameter model embedded in the framework of IFT. The model comprises \emph{a~priori} assumptions for the signal fields that account for the different statistics and correlations of the morphologically different components.
    The diffuse photon flux is assumed to obey multivariate log-normal statistics, where the covariance is described by a power spectrum. The power spectrum is \emph{a~priori} unknown and reconstructed from the data along with the signal. Therefore, hyperpriors on the (logarithmic) power spectra have been introduced, including a spectral smoothness prior \citep{EF11,OSBE12}.
    The point-like photon flux, in contrast, is assumed to factorize spatially in independent inverse-Gamma distributions implying a (regularized) power-law behavior of the amplitudes of the flux.

    An adequate description of the noise properties in terms of a likelihood, here a Poisson distribution, and the incorporation of all instrumental effects into the response operator renders the denoising and deconvolution task possible. The strength of the proposed approach is the performance of the additional decomposition task, which especially exploits the \emph{a~priori} description of diffuse and point-like.
    The model comes down to five scalar parameters, for which all \emph{a~priori} defaults can be motivated, and of which none is driving the inference predominantly.

    We discussed maximum \emph{a~posteriori} (MAP) and Gibbs free energy approaches to solve the inference problem. The derived solutions provide optimal estimators that, in the considered examples, yielded equivalently excellent results. The Gibbs solution slightly outperforms MAP solutions (in terms of the considered L$^2$-residuals) thanks to the full regularization treatment, however, for the price of a computationally more expensive optimization. Which approach is to be preferred in general might depend on the concrete problem at hand and the trade-off between reconstruction precision against computational effort.

    The performance of the D$^3$PO algorithm has been demonstrated in realistic simulations carried out in 1D and 2D. The implementation relies on the \textsc{NIFTy} package \citep{S+13}, which allows for the application regardless of the underlying position space.

    In the 2D application example, a high energy observation of a $32 \times 32 \,\mathrm{arcmin}^2$ patch of a simulated sky with a $0.1 \,\mathrm{arcmin}$ resolution has been analyzed. The D$^3$PO algorithm successfully denoised, deconvolved and decomposed the data image.
    The analysis yielded a detailed reconstruction of the diffuse photon flux and its logarithmic power spectrum, the precise localization of the point sources and accurate determination of their flux intensities, as well as \emph{a~posteriori} estimates of the reconstructed fields.

    The D$^3$PO algorithm should be applicable to a wide range of inference problems appearing in astronomical imaging and related fields. Concrete applications in high energy astrophysics, for example, the analysis of data from the Chandra X-ray observatory or the Fermi $\gamma$-ray space telescope, are currently considered by the authors. In this regard, the public release of the  D$^3$PO code is planned.

%%================================
\section*{Acknowledgments}

    We thank Niels Oppermann, Henrik Junklewitz and two anonymous referees for the insightful discussions and productive comments.

    Furthermore, we thank the DFG Forschergruppe 1254 ``Magnetisation of Interstellar and Intergalactic Media: The Prospects of Low-Frequency Radio Observations'' for travel support in order to present this work at their annual meeting in 2013.

    Some of the results in this publication have been derived using the \textsc{NIFTy} package \citep{S+13}. This research has made use of NASA's Astrophysics Data System.

%%================================
\bibliographystyle{myaa}
%\footnotesize
\bibliography{D3PO.bib}
%\normalsize

\begin{appendix}
%%================================
\section{Point source stacking}
\label{app:stacking}

    In Sec.~\ref{sec:u_prior}, a prior for the point-like signal field has been derived under the assumption that the photon flux of point sources is independent between different pixels and identically inverse-Gamma distributed,
    \begin{align}
        \rho_x^{(u)} &\curvearrowleft \mathcal{I}\left( \rho_x^{(u)}, \beta = \frac{3}{2}, \rho_0\eta \right) \quad \forall x
        ,
    \end{align}
    with the shape and scale parameters, $\beta$ and $\eta$. It can be shown that, for $\beta = \tfrac{3}{2}$, the sum of $N$ such variables still obeys an inverse-Gamma distribution,
    \begin{align}
        \rho_N^{(u)} &= \sum_x^N \rho_x^{(u)}
        \\
        %\sum_x^N \rho_x^{(u)} &\curvearrowleft \mathcal{I}\left( \sum_x^N \rho_x^{(u)}, \beta = \frac{3}{2}, N^2\rho_0\eta \right)
        \rho_N^{(u)} &\curvearrowleft \mathcal{I}\left( \rho_N^{(u)}, \beta = \frac{3}{2}, N^2\rho_0\eta \right)
        .
    \end{align}
    For a proof see \citep{GC01}.

    In the case of $\beta = \tfrac{3}{2}$, the power-law behavior of the prior becomes independent of the discretization of the continuous position space. This means that the slope of the distribution of $\rho_x^{(u)}$ remains unchanged notwithstanding that we refine or coarsen the resolution of the reconstruction. However, the scale parameter $\eta$ needs to be adapted for each resolution; i.e., $\eta \rightarrow N^2\eta$ if $N$ pixels are merged.

%%================================
\section{Covariance \& curvature.}
\label{app:covariance}

    The covariance $\bb{D}$ of a Gaussian $\G(\bb{s}-\bb{m},\bb{D})$ describes the uncertainty associated with the mean $\bb{m}$ of the distribution. It can be computed by second moments or cumulants according to Eq.~\eqref{eq:error}, or in this Gaussian case as the inverse Hessian of the corresponding information Hamiltonian,
    \begin{align}
        \frac{\partial^2 H}{\partial\bb{s}\partial\bb{s}^\T} \bigg|_{\bb{s} = \bb{m}} &= \frac{\partial^2}{\partial\bb{s}\partial\bb{s}^\T} \left( \frac{1}{2} (\bb{s} - \bb{m})^\T \bb{D}^{-1} (\bb{s} - \bb{m}) \right) \bigg|_{\bb{s} = \bb{m}}
        \notag \\
        &= \bb{D}^{-1}
        .
    \end{align}
    In Sec.~\ref{sec:solution}, uncertainty covariances for the diffuse signal field $\bb{s}$ and the point-like signal field $\bb{u}$ have been derived that are here given in closed form.

    The MAP uncertainty covariances introduced in Sec.~\ref{sec:MAP} are approximated by inverse Hessians. According to Eq.~\eqref{eq:D_MAP}, they read
    \begin{align}
        %\frac{\partial^2 H}{\partial s_x \partial s_y} \bigg|_\mathrm{min} &=
        {D_{xy}^{(s)}}^{-1} &\approx
        \left\{ \sum_i \left( 1 - \frac{d_i}{l_i} \right) R_{ix} \e^{m_x^{(s)}} \right\} \delta_{xy}
        \\
        &\quad + \sum_i \frac{d_i}{l_i^2} \left( R_{ix} \e^{m_x^{(s)}} \right) \left( R_{iy} \e^{m_y^{(s)}} \right) + {S_{xy}^\star}^{-1}
        , \notag
    \end{align}
    and
    \begin{align}
        %\frac{\partial^2 H}{\partial u_x \partial u_y} \bigg|_\mathrm{min} &=
        {D_{xy}^{(u)}}^{-1} &\approx
        \left\{ \sum_i \left( 1 - \frac{d_i}{l_i} \right) R_{ix} \e^{m_x^{(u)}} + \eta \: \e^{-m_x^{(u)}} \right\} \delta_{xy}
        \notag \\
        &\quad + \sum_i \frac{d_i}{l_i^2} \left( R_{ix} \e^{m_x^{(u)}} \right) \left( R_{iy} \e^{m_y^{(u)}} \right)
        ,
    \end{align}
    with
    \begin{align}
        l_i = \int\d x \; R_{ix} \left( \e^{m_x^{(s)}} + \e^{m_x^{(u)}} \right)
        .
    \end{align}
    The corresponding covariances derived in the Gibbs approach according to Eq.~\eqref{eq:D_Gibbs}, yield
    \begin{align}
        %\frac{\partial^2 G}{\partial m_x^{(s)} \partial m_y^{(s)}} &=
        {D_{xy}^{(s)}}^{-1} &\approx
        \left\{ \sum_i \left( 1 - \frac{d_i}{l_i} \right) R_{ix} \e^{m_x^{(s)} + \tfrac{1}{2} D_{xx}^{(s)}} \right\} \delta_{xy}
        \\
        &\quad + \sum_i \frac{d_i}{l_i^2} \left( R_{ix} \e^{m_x^{(s)} + \tfrac{1}{2} D_{xx}^{(s)}} \right) \left( R_{iy} \e^{m_y^{(s)} + \tfrac{1}{2} D_{yy}^{(s)}} \right)
        \notag \\
        &\quad + {S_{xy}^\star}^{-1}
        , \notag
    \end{align}
    and
    \begin{align}
        %\frac{\partial^2 G}{\partial m_x^{(u)} \partial m_y^{(u)}} &=
        {D_{xy}^{(u)}}^{-1} &\approx
        \Bigg\{ \sum_i \left( 1 - \frac{d_i}{l_i} \right) R_{ix} \e^{m_x^{(u)} + \tfrac{1}{2} D_{xx}^{(u)}}
        \notag \\
        &\quad \phantom{\Bigg\{} + \eta \: \e^{-m_x^{(u)} + \tfrac{1}{2} D_{xx}^{(u)}} \Bigg\} \delta_{xy}
        \\
        &\quad + \sum_i \frac{d_i}{l_i^2} \left( R_{ix} \e^{m_x^{(u)} + \tfrac{1}{2} D_{xx}^{(u)}} \right) \left( R_{iy} \e^{m_y^{(u)} + \tfrac{1}{2} D_{yy}^{(u)}} \right)
        , \notag
    \end{align}
    with
    \begin{align}
        l_i = \int\d x \; R_{ix} \left( \e^{m_x^{(s)} + \tfrac{1}{2} D_{xx}^{(s)}} + \e^{m_x^{(u)} + \tfrac{1}{2} D_{xx}^{(u)}} \right)
        .
    \end{align}
    They are identical up to the $+ \tfrac{1}{2} D_{xx}$ terms in the exponents. On the one hand, this reinforces the approximations done in Sec.~\ref{sec:Gibbs}. On the other hand, this shows that higher order correction terms might alter the uncertainty covariances further, cf. Eq.~\eqref{eq:G}. The concrete impact of these correction terms is difficult to judge, since they introduce terms involving $D_{xy}$ that couple all elements of $\bb{D}$ in an implicit manner.

    We note that the inverse Hessian describes the curvature of the potential, its interpretation as uncertainty is, strictly speaking, only valid for quadratic potentials. However, in most cases it is a sufficient approximation.

    The Gibbs approach provides an alternative by equating the first derivative of the Gibbs free energy with respect to the covariance with zero. Following Eq.~\eqref{eq:D_Gibbs_2}, the covariances read
    \begin{align}
        {D_{xy}^{(s)}}^{-1}
        &= \left\{ \sum_i \left( 1 - \frac{d_i}{l_i} \right) R_{ix} \e^{m_x^{(s)} + \tfrac{1}{2} D_{xx}^{(s)}} \right\} \delta_{xy}
        \\
        &\quad + {S_{xy}^\star}^{-1}
        ,
    \end{align}
    and
    \begin{align}
        {D_{xy}^{(u)}}^{-1}
        &= \Bigg\{ \sum_i \left( 1 - \frac{d_i}{l_i} \right) R_{ix} \e^{m_x^{(u)} + \tfrac{1}{2} D_{xx}^{(u)}}
        \\
        &\quad \phantom{\Bigg\{} + \eta \: \e^{-m_x^{(u)} + \tfrac{1}{2} D_{xx}^{(u)}} \Bigg\} \delta_{xy}
        .
    \end{align}
    Compared to the above solutions, there is one term missing indicating that they already lack first order corrections. For this reasons, the solutions obtained from the inverse Hessians are used in the D$^3$PO algorithm.

%%================================
\section{Posterior approximation}
\label{app:approximations}
\subsection{Information theoretical measure}
\label{app:KLG}

    If the full posterior $P(\bb{z}|\bb{d})$ of an inference problem is so complex that an analytic handling is infeasible, an approximate posterior $Q$ might be used instead. The fitness of such an approximation can be quantified by an asymmetric measure for which different terminologies appear in the literature.

    First, the Kullback-Leibler divergence,
    \begin{align}
        D_\mathrm{KL}(Q,P) &= \int\D\bb{z} \; Q(\bb{z}|\bb{d}) \log \frac{Q(\bb{z}|\bb{d})}{P(\bb{z}|\bb{d})}
        \\
        &= \left< \log \frac{Q(\bb{z}|\bb{d})}{P(\bb{z}|\bb{d})} \right>_Q
        ,
    \end{align}
    defines mathematically an information theoretical distance, or divergence, which is minimal if a maximal cross information between $P$ and $Q$ exists \citep{KL51}.

    Second, the information entropy,
    \begin{align}
        S_\mathrm{E}(Q,P) &= - \int\D\bb{z} \; P(\bb{z}|\bb{d}) \log \frac{P(\bb{z}|\bb{d})}{Q(\bb{z}|\bb{d})}
        \\
        &= \left< - \log \frac{P(\bb{z}|\bb{d})}{Q(\bb{z}|\bb{d})} \right>_P
        \\
        &= - D_\mathrm{KL}(P,Q)
        , \notag
    \end{align}
    is derived under the maximum entropy principle \citep{J57} from fundamental axioms demanding locality, coordinate invariance and system independence \citet[see e.g.,][]{C08,C11}.

    Third, the (approximate) Gibbs free energy \citep{EW10},
    \begin{align}
        G &= \big< H(\bb{z}|\bb{d}) \big>_Q - S_\mathrm{B}(Q)
        \label{eq:T_1} \\
        &= \big< - \log P(\bb{z}|\bb{d}) \big>_Q - \big< - \log Q(\bb{z}|\bb{d}) \big>_Q
        \\
        &= D_\mathrm{KL}(Q,P)
        , \notag
    \end{align}
    describes the difference between the internal energy $\left< H(\bb{z}|\bb{d}) \right>_Q$ and the Boltzmann-Shannon entropy $S_\mathrm{B}(Q) = S_\mathrm{E}(1,Q)$. The derivation of the Gibbs free energy is based on the principles of thermodynamics\footnote{In Eq.~\eqref{eq:T_1}, a unit temperature is implied, see discussion by \citet{EW10,ISM12,EW12}}.

    The Kullback-Leibler divergence, information entropy, and the Gibbs free energy are equivalent measures that allow one to assess the approximation $Q \approx P$. Alternatively, a parametrized proposal for $Q$ can be pinned down by extremizing the measure of choice with respect to the parameters.

%%--------------------------------
\subsection{Calculus of variations}
\label{app:variation}

    The information theoretical measure can be interpreted as an action to which the principle of least action applies. This concept is the basis for variational Bayesian methods \citep{JGJS99,WW13}, which enable among others the derivation of approximate posterior distributions.

    We suppose that $\bb{z}$ is a set of multiple signal fields, $\bb{z} = \{\bb{z}^{(i)}\}_{i \in \N}$, $\bb{d}$ a given data set, and $P(\bb{z}|\bb{d})$ the posterior of interest. In practice, such a problem is often addressed by a mean field approximation that factorizes the variational posterior $Q$,
    \begin{align}
        P(\bb{z}|\bb{d}) &\approx Q = \prod_i Q_i(\bb{z}^{(i)}|\bb{\mu},\bb{d})
        . \label{eq:meanfield}
    \end{align}
    Here, the mean field $\bb{\mu}$, which mimics the effect of all $\bb{z}^{(i \neq j)}$ onto $\bb{z}^{(j)}$, has been introduced. The approximation in Eq.~\eqref{eq:meanfield} shifts any possible entanglement between the $\bb{z}^{(i)}$ within $P$ into the dependence of $\bb{z}^{(i)}$ on $\bb{\mu}$ within $Q_i$.
    Hence, the mean field $\bb{\mu}$ is well determined by the inference problem at hand, as demonstrated in the subsequent Sect.~\ref{app:about_t}. We note that $\bb{\mu}$ represents effective rather than additional degrees of freedom.

    Following the principle of least action, any variation of the Gibbs free energy must vanish. We consider a variation $\delta_j = \delta/\delta Q_j(\bb{z}^{(j)}|\bb{\mu},\bb{d})$ with respect to one approximate posterior $Q_j(\bb{z}^{(j)}|\bb{\mu},\bb{d})$. It holds,
    \begin{align}
        \frac{\delta Q_i(\tb{z}^{(i)}|\bb{\mu},\bb{d})}{\delta Q_j(\bb{z}^{(j)}|\bb{\mu},\bb{d})} &= \delta_{ij} \; \delta(\bb{z}^{(i)}-\tb{z}^{(j)})
        .
    \end{align}
    Computing the variation of the Gibbs free energy yields

    \begin{widetext}
        \begin{align}
            \delta_j G = 0 &= \frac{\delta}{\delta Q_j(\bb{z}^{(j)}|\bb{\mu},\bb{d})} \; \bigg\{ \big< H(\bb{z}|\bb{d}) \big>_Q - \big< - \log Q \big>_Q \bigg\}
            \label{eq:without_Lagrange} \\
            &= \frac{\delta}{\delta Q_j(\bb{z}^{(j)}|\bb{\mu},\bb{d})} \; \bigg\{ \big< H(\bb{z}|\bb{d}) \big>_Q + \sum_i \big< \log Q_i(\tb{z}^{(i)}|\bb{\mu},\bb{d}) \big>_{Q_i} \bigg\}
            \\
            &= \frac{\delta}{\delta Q_j(\bb{z}^{(j)}|\bb{\mu},\bb{d})} \; \int \D\tb{z}^{(j)} \; Q_j(\tb{z}^{(j)}|\bb{\mu},\bb{d}) \; \bigg\{ \Big< H(\bb{z}|\bb{d}) \Big>_{\prod Q_{i \neq j}} + \log Q_j(\tb{z}^{(j)}|\bb{\mu},\bb{d}) \bigg\} + \underbrace{\frac{\delta}{\delta Q_j(\bb{z}^{(j)}|\bb{\mu},\bb{d})} \sum_{i \neq j} \dots}_{=0}
            \nonumber \\
            &= \int \D\tb{z}^{(j)} \; \delta(\bb{z}^{(i)}-\tb{z}^{(j)}) \; \bigg\{ \Big< H(\bb{z}|\bb{d}) \Big>_{\prod Q_{i \neq j}} + \log Q_j(\tb{z}^{(j)}|\bb{\mu},\bb{d}) + 1 \bigg\}
            \nonumber \\
            &= \left< H(\bb{z}|\bb{d}) \Big|_{\bb{z}^{(j)}} \right>_{\prod Q_{i \neq j}} + \log Q_j(\bb{z}^{(j)}|\bb{\mu},\bb{d}) + \mathrm{const.}
            \label{eq:variation}
        \end{align}
    \end{widetext}
    This defines a solution for the approximate posterior $Q_j$, where the constant term in Eq.~\eqref{eq:variation} ensures the correct normalization\footnote{The normalization could be included by usage of Lagrange multipliers; i.e., by adding a term $\sum_i \lambda_i \big( 1 - \int\D\bb{z}^{(i)} \; Q_i(\bb{z}^{(i)}|\bb{\mu},\bb{d}) \big)$ to the Gibbs free energy in Eq.~\eqref{eq:without_Lagrange}.} of $Q_j$,
    \begin{align}
        Q_j(\bb{z}^{(j)}|\bb{\mu},\bb{d}) &\propto \exp \left( - \left< H(\bb{z}|\bb{d}) \Big|_{\bb{z}^{(j)}} \right>_{\prod Q_{i \neq j}} \right)
        . \label{eq:variational_Q}
    \end{align}
    Although the parts $\bb{z}^{(i \neq j)}$ are integrated out, Eq.~\eqref{eq:variational_Q} is no marginalization since the integration is performed on the level of the (negative) logarithm of a probability distribution. The success of the mean field approach might be that this integration is often more well-behaved in comparison to the corresponding marginalization. However, the resulting equations for the $Q_i$ depend on each other, and thus need to be solved self-consistently.

    A maximum \emph{a~posteriori} solution for $\bb{z}^{(j)}$ can then be found by minimizing an effective Hamiltonian,
    \begin{align}
        \underset{\bb{z}^{(j)}}{\mathrm{argmax}} \; P(\bb{z}|\bb{d}) &= \underset{\bb{z}^{(j)}}{\mathrm{argmin}} \; H(\bb{z}|\bb{d})
        \\
        &\approx \underset{\bb{z}^{(j)}}{\mathrm{argmin}} \; \left< H(\bb{z}|\bb{d}) \Big|_{\bb{z}^{(j)}} \right>_{\prod Q_{i \neq j}}
        .
    \end{align}
    Since the posterior is approximated by a product, the Hamiltonian is approximated by a sum, and each summand depends on solely one variable in the partition of the latent variable $\bb{z}$.

%%--------------------------------
\subsection{Example}
\label{app:about_t}

    \begin{figure}[t!]
        \centering
        \begin{tikzpicture}
            [c/.style={circle,minimum size=2em,text centered,thin},
             r/.style={rectangle,minimum size=2em,text centered,thin},
             v/.style={->,shorten >=1pt,>=stealth,thick}]
            \node(aa)at(-4.25,3.6)[]{(a)};
            \node(m1)at(-2.25,3)[r,text width=4em,draw]{model};
            \node(xx)at(-2.25,2)[]{};
            \node(t1)at(-3.25,1)[c,draw]{$\bb{\tau}$};
            \node(s1)at(-1.25,1)[c,draw]{$\bb{s}$};
            \node(d1)at(-2.25,0)[r,draw]{$\bb{d}$};
            \draw[v](m1)--(xx.center)--(t1);
            \draw[v](t1)--(s1);
            \draw[v](s1)--(d1);
            \node(bb)at(0.25,3.6)[]{(b)};
            \node(m2)at(2.25,3)[r,text width=4em,draw]{model};
            \node(mu)at(2.25,2)[c,draw,dashed]{$\bb{\mu}$};
            \node(t2)at(1.25,1)[c,draw]{$\bb{\tau}$};
            \node(s2)at(3.25,1)[c,draw]{$\bb{s}$};
            \node(d2)at(2.25,0)[r,draw]{$\bb{d}$};
            \draw[v](m2)--(mu);
            \draw[v](mu)--(t2);
            \draw[v](mu)--(s2);
            \draw[v](t2)--(d2);
            \draw[v](s2)--(d2);
        \end{tikzpicture}
        \flushleft
        \caption{Graphical model for the variational method applied to the example posterior in Eq.~\eqref{eq:example}. Panel (a) shows the graphical model without, and panel (b) with the mean field $\bb{\mu}$.}
        \label{fig:mu}
    \end{figure}
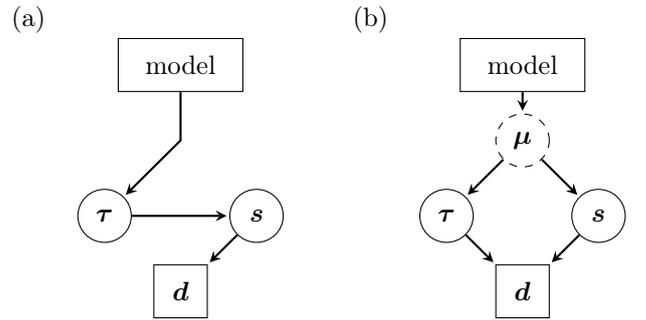

    In this section, the variational method is demonstrated with an exemplary posterior of the following form,
    \begin{align}
        P(\bb{s},\bb{\tau}|\bb{d}) &= \frac{P(\bb{d}|\bb{s})}{P(\bb{d})} \; P(\bb{s}|\bb{\tau}) \; P(\bb{\tau})
        \label{eq:example} \\
        &= \frac{P(\bb{d}|\bb{s})}{P(\bb{d})} \; \G(\bb{s},\bb{S}) \; P_\mathrm{un}(\bb{\tau}|\bb{\alpha},\bb{q}) \; P_\mathrm{sm}(\bb{\tau}|\bb{\sigma})
        ,
    \end{align}
    where $P(\bb{d}|\bb{s})$ stands for an arbitrary likelihood describing how likely the data $\bb{d}$ can be measured from a signal $\bb{s}$, and $\bb{S} = \sum_k \e^{\tau_k} \bb{S}_k$ for a parametrization of the signal covariance. This posterior is equivalent to the one derived in Sec.~\ref{sec:problem} in order to find a solution for the logarithmic power spectrum $\bb{\tau}$. Here, any explicit dependence on the point-like signal field $\bb{u}$ is veiled in favor of clarity.

    The corresponding Hamiltonian reads
    \begin{align}
        H(\bb{s},\bb{\tau}|\bb{d}) &= - \log P(\bb{s},\bb{\tau}|\bb{d})
        \\
        &= H_0 + \frac{1}{2} \sum_k \left( \varrho_k \tau_k + \tr\left[ \bb{s}\bb{s}^\T \bb{S}_k^{-1} \right] \e^{-\tau_k} \right)
        \label{eq:H_st} \\
        &\quad + (\bb{\alpha} - \bb{1})^\T \bb{\tau} + \bb{q}^\T \e^\bb{-\tau} + \frac{1}{2} \bb{\tau}^\T \bb{T} \bb{\tau}
        , \notag
    \end{align}
    where $\varrho_k = \tr\left[ \bb{S}_k {\bb{S}_k}^{-1} \right]$ and all terms constant in $\bb{\tau}$, including the likelihood $P(\bb{d}|\bb{s})$, have been absorbed into $H_0$.

    For an arbitrary likelihood it might not be possible to marginalize the posterior over $\bb{s}$ analytically. However, an integration of the Hamiltonian over $\bb{s}$ might be feasible since the only relevant term is quadratic in $\bb{s}$. As, on the one hand, the prior $P(\bb{s}|\bb{\tau})$ is Gaussian and, on the other hand, a posterior mean $\bb{m}$ and covariance $\bb{D}$ for the signal field $\bb{s}$ suffice, cf. Eq.~\eqref{eq:map} and \eqref{eq:error}, we assume a Gaussian approximation for $Q_s$; i.e., $Q_s = \G(\bb{s}-\bb{m},\bb{D})$.

    We now introduce a mean field approximation, denoted by $\bb{\mu}$, by changing the causal structure as depicted in Fig.~\ref{fig:mu}. With the consequential approximation of the posterior,
    \begin{align}
        P(\bb{s},\bb{\tau}|\bb{d}) &\approx \G(\bb{s}-\bb{m},\bb{D}) \; Q_\tau(\bb{\tau}|\bb{\mu},\bb{d})
        ,
    \end{align}
    we can calculate the effective Hamiltonian for $\bb{\tau}$ as
    \begin{align}
        \left< H(\bb{s},\bb{\tau}|\bb{d}) \Big|_{\bb{\tau}} \right>_{Q_s} &= H_0  + \bb{\gamma}^\T \bb{\tau} + \frac{1}{2} \bb{\tau}^\T \bb{T} \bb{\tau} + \bb{q}^\T \e^\bb{-\tau}
        \\
        &\quad + \frac{1}{2} \sum_k \tr\left[ \left< \bb{s}\bb{s}^\T \right>_{Q_s} \bb{S}_k^{-1} \right] \e^{-\tau_k}
        \notag \\
        &= H_0  + \bb{\gamma}^\T \bb{\tau} + \frac{1}{2} \bb{\tau}^\T \bb{T} \bb{\tau} + \bb{q}^\T \e^\bb{-\tau}
        \label{eq:H_t} \\
        &\quad + \frac{1}{2} \sum_k \tr\Big[ \left( \bb{m}\bb{m}^\T + \bb{D} \right) \bb{S}_k^{-1} \Big] \e^{-\tau_k}
        , \notag
    \end{align}
    where $\bb{\gamma} = (\bb{\alpha} - \bb{1}) + \frac{1}{2} \bb{\varrho}$.

    The nature of the mean field $\bb{\mu}$ can be derived from the coupling term in Eq.~\eqref{eq:H_st} that ensures an information flow between $\bb{s}$ and $\bb{\tau}$,
    \begin{align}
        \bb{\mu} &= \begin{pmatrix} \left< \tr\left[ \bb{s}\bb{s}^\T \bb{S}_k^{-1} \right] \right>_{Q_s} \\ \left< \sum_k \e^{-\tau_k} \bb{S}_k^{-1}  \right>_{Q_\tau} \end{pmatrix} = \begin{pmatrix} \tr\left[ \left( \bb{m}\bb{m}^\T + \bb{D} \right) \bb{S}_k^{-1} \right] \\ \left< \bb{S}^{-1} \right>_{Q_\tau} \end{pmatrix}
    \end{align}
    Hence, the mean field effect on $\tau_k$ is given by the above trace, and the mean field effect on $\bb{s}$ is described by $\left< \bb{S}^{-1} \right>_{Q_\tau}$.

    Extremizing Eq.~\eqref{eq:H_t} yields
    \begin{align}
        \e^{\bb{\tau}} &= \frac{\bb{q} + \frac{1}{2} \left( \tr\left[ \left( \bb{m}\bb{m}^\T + \bb{D} \right) \bb{S}_k^{-1} \right] \right)_k}{\bb{\gamma} + \bb{T} \bb{\tau}}
        .
    \end{align}
    This formula is in agreement with the critical filter formula \citep{EF11,OSBE12}. In case a Gaussian likelihood and no smoothness prior is assumed, it is the exact maximum of the true posterior with respect to the (logarithmic) power spectrum.

\end{appendix}

\setcounter{table}{1}
\begin{table*}
    \caption{Overview of the relative residual error in the photon flux reconstructions for a MAP-$\delta$ approach with varying model parameters $\sigma$, $\beta$, and $\eta$. The parameters $\alpha$ and $q$ were fixed. The best and worst residuals are printed in bold face.}
    \centering
    \begin{tabular}{|ll|lllll|}
\hline
$\alpha = 1$ & $q = 10^{-12}$ & \multicolumn{1}{|c}{$\sigma = 1$} & \multicolumn{1}{c}{$\sigma = 10$} & \multicolumn{1}{c}{$\sigma = 100$} & \multicolumn{1}{c}{$\sigma = 1000$} & \multicolumn{1}{c|}{$\sigma \rightarrow \infty$} \\
\hline
\hline
\multirow{2}{*}{$\beta = 1$}  & \multirow{2}{*}{$\eta = 10^{-6}$} & $\phantom{^u}\epsilon^{(s)} = 0.06710$ & $\phantom{^u}\epsilon^{(s)} = 0.05406$ & $\phantom{^u}\epsilon^{(s)} = 0.05323$ & $\phantom{^u}\epsilon^{(s)} = 0.05383$ & $\phantom{^u}\epsilon^{(s)} = 0.05359$ \\
&& $\phantom{^s}\epsilon^{(u)} = 0.02000$ & $\phantom{^s}\epsilon^{(u)} = 0.01941$ & $\phantom{^s}\epsilon^{(u)} = 0.01602$ & $\phantom{^s}\epsilon^{(u)} = 0.01946$ & $\phantom{^s}\epsilon^{(u)} = 0.01898$ \\
\multirow{2}{*}{$\beta = \tfrac{5}{4}$}  & \multirow{2}{*}{$\eta = 10^{-6}$} & $\phantom{^u}\epsilon^{(s)} = 0.02874$ & $\phantom{^u}\epsilon^{(s)} = 0.01929$ & $\phantom{^u}\epsilon^{(s)} = 0.01974$ & $\phantom{^u}\epsilon^{(s)} = 0.02096$ & $\phantom{^u}\epsilon^{(s)} = 0.01991$ \\
&& $\phantom{^s}\epsilon^{(u)} = 0.01207$ & $\phantom{^s}\epsilon^{(u)} = 0.01102$ & $\phantom{^s}\epsilon^{(u)} = \bb{0.01090}$ & $\phantom{^s}\epsilon^{(u)} = 0.01123$ & $\phantom{^s}\epsilon^{(u)} = 0.01104$ \\
\multirow{2}{*}{$\beta = \tfrac{3}{2}$}  & \multirow{2}{*}{$\eta = 10^{-6}$} & $\phantom{^u}\epsilon^{(s)} = 0.05890$ & $\phantom{^u}\epsilon^{(s)} = 0.02237$ & $\phantom{^u}\epsilon^{(s)} = 0.02318$ & $\phantom{^u}\epsilon^{(s)} = 0.02238$ & $\phantom{^u}\epsilon^{(s)} = 0.02344$ \\
&& $\phantom{^s}\epsilon^{(u)} = 0.02741$ & $\phantom{^s}\epsilon^{(u)} = 0.01343$ & $\phantom{^s}\epsilon^{(u)} = 0.01346$ & $\phantom{^s}\epsilon^{(u)} = 0.01342$ & $\phantom{^s}\epsilon^{(u)} = 0.01351$ \\
\multirow{2}{*}{$\beta = \tfrac{7}{4}$}  & \multirow{2}{*}{$\eta = 10^{-6}$} & $\phantom{^u}\epsilon^{(s)} = 0.10864$ & $\phantom{^u}\epsilon^{(s)} = 0.04304$ & $\phantom{^u}\epsilon^{(s)} = 0.03234$ & $\phantom{^u}\epsilon^{(s)} = 0.03248$ & $\phantom{^u}\epsilon^{(s)} = 0.03263$ \\
&& $\phantom{^s}\epsilon^{(u)} = 0.04840$ & $\phantom{^s}\epsilon^{(u)} = 0.02767$ & $\phantom{^s}\epsilon^{(u)} = 0.02142$ & $\phantom{^s}\epsilon^{(u)} = 0.02143$ & $\phantom{^s}\epsilon^{(u)} = 0.02167$ \\
\multirow{2}{*}{$\beta = 2$}  & \multirow{2}{*}{$\eta = 10^{-6}$} & $\phantom{^u}\epsilon^{(s)} = 0.11870$ & $\phantom{^u}\epsilon^{(s)} = 0.04614$ & $\phantom{^u}\epsilon^{(s)} = 0.04527$ & $\phantom{^u}\epsilon^{(s)} = 0.04522$ & $\phantom{^u}\epsilon^{(s)} = 0.04500$ \\
&& $\phantom{^s}\epsilon^{(u)} = 0.05360$ & $\phantom{^s}\epsilon^{(u)} = 0.02926$ & $\phantom{^s}\epsilon^{(u)} = 0.02924$ & $\phantom{^s}\epsilon^{(u)} = 0.02926$ & $\phantom{^s}\epsilon^{(u)} = 0.02915$ \\
\hline
\multirow{2}{*}{$\beta = 1$}  & \multirow{2}{*}{$\eta = 10^{-4}$} & $\phantom{^u}\epsilon^{(s)} = 0.06660$ & $\phantom{^u}\epsilon^{(s)} = 0.05474$ & $\phantom{^u}\epsilon^{(s)} = 0.05377$ & $\phantom{^u}\epsilon^{(s)} = 0.05474$ & $\phantom{^u}\epsilon^{(s)} = 0.05423$ \\
&& $\phantom{^s}\epsilon^{(u)} = 0.02157$ & $\phantom{^s}\epsilon^{(u)} = 0.01903$ & $\phantom{^s}\epsilon^{(u)} = 0.01657$ & $\phantom{^s}\epsilon^{(u)} = 0.01986$ & $\phantom{^s}\epsilon^{(u)} = 0.02055$ \\
\multirow{2}{*}{$\beta = \tfrac{5}{4}$}  & \multirow{2}{*}{$\eta = 10^{-4}$} & $\phantom{^u}\epsilon^{(s)} = 0.02874$ & $\phantom{^u}\epsilon^{(s)} = \bb{0.01929}$ & $\phantom{^u}\epsilon^{(s)} = 0.01974$ & $\phantom{^u}\epsilon^{(s)} = 0.02096$ & $\phantom{^u}\epsilon^{(s)} = 0.01991$ \\
&& $\phantom{^s}\epsilon^{(u)} = 0.01207$ & $\phantom{^s}\epsilon^{(u)} = 0.01100$ & $\phantom{^s}\epsilon^{(u)} = 0.01103$ & $\phantom{^s}\epsilon^{(u)} = 0.01123$ & $\phantom{^s}\epsilon^{(u)} = 0.01102$ \\
\multirow{2}{*}{$\beta = \tfrac{3}{2}$}  & \multirow{2}{*}{$\eta = 10^{-4}$} & $\phantom{^u}\epsilon^{(s)} = 0.05890$ & $\phantom{^u}\epsilon^{(s)} = 0.02237$ & $\phantom{^u}\epsilon^{(s)} = 0.02318$ & $\phantom{^u}\epsilon^{(s)} = 0.02238$ & $\phantom{^u}\epsilon^{(s)} = 0.02344$ \\
&& $\phantom{^s}\epsilon^{(u)} = 0.02743$ & $\phantom{^s}\epsilon^{(u)} = 0.01343$ & $\phantom{^s}\epsilon^{(u)} = 0.01346$ & $\phantom{^s}\epsilon^{(u)} = 0.01340$ & $\phantom{^s}\epsilon^{(u)} = 0.01352$ \\
\multirow{2}{*}{$\beta = \tfrac{7}{4}$}  & \multirow{2}{*}{$\eta = 10^{-4}$} & $\phantom{^u}\epsilon^{(s)} = 0.10864$ & $\phantom{^u}\epsilon^{(s)} = 0.04304$ & $\phantom{^u}\epsilon^{(s)} = 0.03234$ & $\phantom{^u}\epsilon^{(s)} = 0.03248$ & $\phantom{^u}\epsilon^{(s)} = 0.03263$ \\
&& $\phantom{^s}\epsilon^{(u)} = 0.04840$ & $\phantom{^s}\epsilon^{(u)} = 0.02766$ & $\phantom{^s}\epsilon^{(u)} = 0.02145$ & $\phantom{^s}\epsilon^{(u)} = 0.02142$ & $\phantom{^s}\epsilon^{(u)} = 0.02166$ \\
\multirow{2}{*}{$\beta = 2$}  & \multirow{2}{*}{$\eta = 10^{-4}$} & $\phantom{^u}\epsilon^{(s)} = 0.11870$ & $\phantom{^u}\epsilon^{(s)} = 0.04614$ & $\phantom{^u}\epsilon^{(s)} = 0.04527$ & $\phantom{^u}\epsilon^{(s)} = 0.04522$ & $\phantom{^u}\epsilon^{(s)} = 0.04500$ \\
&& $\phantom{^s}\epsilon^{(u)} = 0.05358$ & $\phantom{^s}\epsilon^{(u)} = 0.02926$ & $\phantom{^s}\epsilon^{(u)} = 0.02926$ & $\phantom{^s}\epsilon^{(u)} = 0.02927$ & $\phantom{^s}\epsilon^{(u)} = 0.02916$ \\
\hline
\multirow{2}{*}{$\beta = 1$}  & \multirow{2}{*}{$\eta = 10^{-2}$} & $\phantom{^u}\epsilon^{(s)} = 0.07271$ & $\phantom{^u}\epsilon^{(s)} = 0.06209$ & $\phantom{^u}\epsilon^{(s)} = 0.06192$ & $\phantom{^u}\epsilon^{(s)} = 0.06291$ & $\phantom{^u}\epsilon^{(s)} = 0.06265$ \\
&& $\phantom{^s}\epsilon^{(u)} = 0.02252$ & $\phantom{^s}\epsilon^{(u)} = 0.02047$ & $\phantom{^s}\epsilon^{(u)} = 0.02109$ & $\phantom{^s}\epsilon^{(u)} = 0.01764$ & $\phantom{^s}\epsilon^{(u)} = 0.02068$ \\
\multirow{2}{*}{$\beta = \tfrac{5}{4}$}  & \multirow{2}{*}{$\eta = 10^{-2}$} & $\phantom{^u}\epsilon^{(s)} = 0.02335$ & $\phantom{^u}\epsilon^{(s)} = 0.01934$ & $\phantom{^u}\epsilon^{(s)} = 0.02042$ & $\phantom{^u}\epsilon^{(s)} = 0.01999$ & $\phantom{^u}\epsilon^{(s)} = 0.01930$ \\
&& $\phantom{^s}\epsilon^{(u)} = 0.01139$ & $\phantom{^s}\epsilon^{(u)} = 0.01112$ & $\phantom{^s}\epsilon^{(u)} = 0.01097$ & $\phantom{^s}\epsilon^{(u)} = 0.01124$ & $\phantom{^s}\epsilon^{(u)} = 0.01102$ \\
\multirow{2}{*}{$\beta = \tfrac{3}{2}$}  & \multirow{2}{*}{$\eta = 10^{-2}$} & $\phantom{^u}\epsilon^{(s)} = 0.05999$ & $\phantom{^u}\epsilon^{(s)} = 0.02227$ & $\phantom{^u}\epsilon^{(s)} = 0.02347$ & $\phantom{^u}\epsilon^{(s)} = 0.02266$ & $\phantom{^u}\epsilon^{(s)} = 0.02274$ \\
&& $\phantom{^s}\epsilon^{(u)} = 0.02745$ & $\phantom{^s}\epsilon^{(u)} = 0.01341$ & $\phantom{^s}\epsilon^{(u)} = 0.01356$ & $\phantom{^s}\epsilon^{(u)} = 0.01332$ & $\phantom{^s}\epsilon^{(u)} = 0.01351$ \\
\multirow{2}{*}{$\beta = \tfrac{7}{4}$}  & \multirow{2}{*}{$\eta = 10^{-2}$} & $\phantom{^u}\epsilon^{(s)} = 0.10715$ & $\phantom{^u}\epsilon^{(s)} = 0.04304$ & $\phantom{^u}\epsilon^{(s)} = 0.03254$ & $\phantom{^u}\epsilon^{(s)} = 0.03264$ & $\phantom{^u}\epsilon^{(s)} = 0.03258$ \\
&& $\phantom{^s}\epsilon^{(u)} = 0.04833$ & $\phantom{^s}\epsilon^{(u)} = 0.02766$ & $\phantom{^s}\epsilon^{(u)} = 0.02140$ & $\phantom{^s}\epsilon^{(u)} = 0.02144$ & $\phantom{^s}\epsilon^{(u)} = 0.02163$ \\
\multirow{2}{*}{$\beta = 2$}  & \multirow{2}{*}{$\eta = 10^{-2}$} & $\phantom{^u}\epsilon^{(s)} = 0.12496$ & $\phantom{^u}\epsilon^{(s)} = 0.04614$ & $\phantom{^u}\epsilon^{(s)} = 0.04497$ & $\phantom{^u}\epsilon^{(s)} = 0.04528$ & $\phantom{^u}\epsilon^{(s)} = 0.04500$ \\
&& $\phantom{^s}\epsilon^{(u)} = \bb{0.05361}$ & $\phantom{^s}\epsilon^{(u)} = 0.02927$ & $\phantom{^s}\epsilon^{(u)} = 0.02915$ & $\phantom{^s}\epsilon^{(u)} = 0.02914$ & $\phantom{^s}\epsilon^{(u)} = 0.02915$ \\
\hline
\multirow{2}{*}{$\beta = 1$}  & \multirow{2}{*}{$\eta = 1$} & $\phantom{^u}\epsilon^{(s)} = 0.15328$ & $\phantom{^u}\epsilon^{(s)} = 0.14544$ & $\phantom{^u}\epsilon^{(s)} = 0.14138$ & $\phantom{^u}\epsilon^{(s)} = 0.14181$ & $\phantom{^u}\epsilon^{(s)} = 0.14185$ \\
&& $\phantom{^s}\epsilon^{(u)} = 0.03250$ & $\phantom{^s}\epsilon^{(u)} = 0.03291$ & $\phantom{^s}\epsilon^{(u)} = 0.02905$ & $\phantom{^s}\epsilon^{(u)} = 0.03087$ & $\phantom{^s}\epsilon^{(u)} = 0.02876$ \\
\multirow{2}{*}{$\beta = \tfrac{5}{4}$}  & \multirow{2}{*}{$\eta = 1$} & $\phantom{^u}\epsilon^{(s)} = \bb{0.15473}$ & $\phantom{^u}\epsilon^{(s)} = 0.14406$ & $\phantom{^u}\epsilon^{(s)} = 0.14357$ & $\phantom{^u}\epsilon^{(s)} = 0.14465$ & $\phantom{^u}\epsilon^{(s)} = 0.13964$ \\
&& $\phantom{^s}\epsilon^{(u)} = 0.03217$ & $\phantom{^s}\epsilon^{(u)} = 0.03166$ & $\phantom{^s}\epsilon^{(u)} = 0.03089$ & $\phantom{^s}\epsilon^{(u)} = 0.03101$ & $\phantom{^s}\epsilon^{(u)} = 0.03160$ \\
\multirow{2}{*}{$\beta = \tfrac{3}{2}$}  & \multirow{2}{*}{$\eta = 1$} & $\phantom{^u}\epsilon^{(s)} = 0.15360$ & $\phantom{^u}\epsilon^{(s)} = 0.14216$ & $\phantom{^u}\epsilon^{(s)} = 0.14248$ & $\phantom{^u}\epsilon^{(s)} = 0.14208$ & $\phantom{^u}\epsilon^{(s)} = 0.14233$ \\
&& $\phantom{^s}\epsilon^{(u)} = 0.03262$ & $\phantom{^s}\epsilon^{(u)} = 0.03063$ & $\phantom{^s}\epsilon^{(u)} = 0.02534$ & $\phantom{^s}\epsilon^{(u)} = 0.02872$ & $\phantom{^s}\epsilon^{(u)} = 0.03095$ \\
\multirow{2}{*}{$\beta = \tfrac{7}{4}$}  & \multirow{2}{*}{$\eta = 1$} & $\phantom{^u}\epsilon^{(s)} = 0.15206$ & $\phantom{^u}\epsilon^{(s)} = 0.14156$ & $\phantom{^u}\epsilon^{(s)} = 0.13772$ & $\phantom{^u}\epsilon^{(s)} = 0.14160$ & $\phantom{^u}\epsilon^{(s)} = 0.14390$ \\
&& $\phantom{^s}\epsilon^{(u)} = 0.03262$ & $\phantom{^s}\epsilon^{(u)} = 0.03065$ & $\phantom{^s}\epsilon^{(u)} = 0.03174$ & $\phantom{^s}\epsilon^{(u)} = 0.03141$ & $\phantom{^s}\epsilon^{(u)} = 0.03178$ \\
\multirow{2}{*}{$\beta = 2$}  & \multirow{2}{*}{$\eta = 1$} & $\phantom{^u}\epsilon^{(s)} = 0.06421$ & $\phantom{^u}\epsilon^{(s)} = 0.05479$ & $\phantom{^u}\epsilon^{(s)} = 0.05365$ & $\phantom{^u}\epsilon^{(s)} = 0.05499$ & $\phantom{^u}\epsilon^{(s)} = 0.05429$ \\
&& $\phantom{^s}\epsilon^{(u)} = 0.02043$ & $\phantom{^s}\epsilon^{(u)} = 0.01966$ & $\phantom{^s}\epsilon^{(u)} = 0.01676$ & $\phantom{^s}\epsilon^{(u)} = 0.02070$ & $\phantom{^s}\epsilon^{(u)} = 0.01996$ \\
\hline
\end{tabular}

    \label{tab:runs}
\end{table*}

\end{document}